\documentclass[11pt,preprint,trackchanges]{aastex63}

\usepackage{graphics,graphicx}
\usepackage{natbib}
\usepackage{float}
\usepackage{xcolor}
\def\apjl{\rm ApJL}
\def\apjs{\rm ApJS}

\def\aap{\rm AAP}

\newcommand{\gtorder}{\mathrel{\raise.3ex\hbox{$>$}\mkern-14mu
            \lower0.6ex\hbox{$\sim$}}}
\newcommand{\ltorder}{\mathrel{\raise.3ex\hbox{$<$}\mkern-14mu
            \lower0.6ex\hbox{$\sim$}}}

\shorttitle{The High-Density Equation of State}
\shortauthors{Miller, Chirenti, and Lamb}

\begin{document}

\title{Constraining the equation of state of high-density cold matter using nuclear and astronomical measurements
}

\correspondingauthor{M. C. Miller}
\email{miller@astro.umd.edu}

\author{%
    \parbox{\linewidth}{\centering
M.~C.~Miller$^1$,
C. Chirenti$^2$,
F.~K.~Lamb$^{3,4}$
  }%
}%

\affil{{Department of Astronomy and Joint Space-Science Institute, University of Maryland, College Park, MD 20742-2421, USA}\\
{$^2$}{Centro de Matem\'{a}tica, Computa\c{c}{\~a}o e Cogni\c{c}{\~a}o, UFABC, 09210-170 Santo Andr\'{e}-SP, Brazil}\\
{$^3$}{Center for Theoretical Astrophysics and Department of Physics, University of Illinois at Urbana-Champaign, 1110 West Green Street, Urbana, IL 61801-3080, USA}\\
{$^4$}{Department of Astronomy, University of Illinois at Urbana-Champaign, 1002 West Green Street, Urbana, IL 61801-3074, USA}\\
}

\begin{abstract}

The increasing richness of data related to cold dense matter, from laboratory experiments to neutron-star observations, requires a framework for constraining the properties of such matter that makes use of all relevant information. Here, we present a rigorous but practical Bayesian approach that can include diverse evidence, such as nuclear data and the inferred masses, radii, tidal deformabilities, moments of inertia, and gravitational binding energies of neutron stars. We emphasize that the full posterior probability distributions of measurements should be used rather than, as is common, imposing a cut on the maximum mass or other quantities.  Our method can be used with any parameterization of the equation of state (EOS). We use both a spectral parameterization and a piecewise polytropic parameterization with variable transition densities to illustrate the implications of current measurements and show how future measurements in many domains could improve our understanding of cold catalyzed matter. We find that different types of measurements will play distinct roles in constraining the EOS in different density ranges.  For example, better symmetry energy measurements will have a major influence on our understanding of matter somewhat below nuclear saturation density but little influence above that density.  In contrast, precise radius measurements or multiple tidal deformability measurements of the quality of those from GW170817 or better will improve our knowledge of the EOS over a broader density range.

\end{abstract}


\section{Introduction}
\label{sec:introduction}

Several recent or upcoming astronomical measurements have or will have important implications for our understanding of the properties of the cold, catalyzed matter in the cores of neutron stars.  Chief among them are the measurements of the binary tidal deformability from the gravitational-wave event GW170817 \citep{2017PhRvL.119p1101A,2019PhRvX...9a1001A} and the expected measurements of neutron-star radii and masses using the {\it Neutron Star Interior Composition Explorer} ({\it NICER}; \citealt{2016SPIE.9905E..1HG}).  This information, combined with nuclear data and other astronomical constraints, such as the high measured masses of a few neutron stars \citep{2010Natur.467.1081D,2013Sci...340..448A,2019arXiv190406759T}, opens up new opportunities to constrain the equation of state (EOS) of cold high-density matter.

Here, we present a rigorous and practical Bayesian procedure that can be used to combine information from different types of nuclear measurements and observations of neutron star systems to constrain the EOS of high-density cold matter. Our procedure can also be used to constrain other properties of neutron stars. For example, data on the cooling of neutron stars could be used to constrain the composition of the interior of neutron stars (see Potekhin et al. 2015 and Wijnands et al. 2017 for recent reviews), providing information that would be complementary to constraints on the EOS.  Here, however, we focus only on constraints on the EOS.  In Section~\ref{sec:statistics} we discuss our general statistical methodology.  In Section~\ref{sec:specific} we discuss the use of particular types of data, such as the highest measured masses of neutron stars and tidal deformabilities from individual events.  In Section~\ref{sec:previous} we compare our methodology to previous work on constraining the high-density EOS.   In Section~\ref{sec:results} we discuss our assumed priors and present the results for our illustrative EOS models. We summarize our conclusions in Section 6. 

\section{Statistical approach}
\label{sec:statistics}

In this section we describe our methodology. We suppose that different types of observations have been made of a set of neutron stars and that we are considering one or more parameterized models of the EOS of neutron-star matter. How should we analyze these observations to correctly use all the available information to obtain estimates of the posterior probability densities of the parameters in these EOS models?

We assume that $n$ neutron stars have been observed and denote a property of star $i$ by a subscript $i$ on the symbol for that property. The observations can be of very different types, e.g., separate measurements of different stars could inform us about their masses, or masses and radii, or moments of inertia, or tidal deformabilities.  Our notation is:
\begin{equation}
\begin{array}{rl}
{\vec\alpha}&\qquad{\rm Equation~of~state~parameters}\\
{\rho_{c,i}}&\qquad{\rm Central~density~of~star}~i\\
{\vec\beta}_i&\qquad{\rm Other~parameters~fixed~for~star}~i\\
{\vec\gamma}_{i,j}&\qquad{\rm Parameters~that~could~vary~between~measurements}~j~{\rm of~star}~i\\
q(\ldots)&\qquad{\rm Prior~probability~density}\\
P(\ldots)&\qquad{\rm Posterior~probability~density}\\
{\cal L}(\ldots)&\qquad{\rm Likelihood~of~the~data~given~a~model~with~associated~parameter~values.}\\
\end{array}
\end{equation}
Note that given ${\vec\alpha}$ and a rotation rate, $\rho_{c,i}$ determines the mass $M_i$ of star $i$.  By assumption, the true value of ${\vec\alpha}$ is the same for all neutron stars, and the true value of $\rho_{c,i}$ is fixed for a given star (and thus does not vary with the measurement $j$), but can vary from one star to another.  Examples of other parameters that are fixed for a given star are the observer inclination and distance to the star; those parameters can, of course, vary from one star to another.  The other parameters ${\vec\gamma}_{i,j}$ (such as the surface emission pattern during a thermonuclear burst) can vary from one measurement to the next of a single star and can vary from one star to another.  The measurements could be of entirely distinct types.

We are interested in the posterior probability density $P({\vec\alpha})$.  We obtain this by marginalizing the full posterior probability density $P({\vec\alpha},\rho_{c,i},{\vec\beta}_i,{\vec\gamma}_{i,j})\propto q({\vec\alpha},\rho_{c,i},{\vec\beta}_i,{\vec\gamma}_{i,j}){\cal L}({\vec\alpha},\rho_{c,i},{\vec\beta}_i,{\vec\gamma}_{i,j})$ over the nuisance parameters (i.e., the parameters that do not depend directly on the EOS) $\rho_{c,i}$, ${\vec\beta}_i$, and ${\vec\gamma}_{i,j}$:
\begin{equation}
P({\vec\alpha})\propto \int q({\vec\alpha},\rho_{c,i},{\vec\beta}_i,{\vec\gamma}_{i,j}){\cal L}({\vec\alpha},\rho_{c,i},{\vec\beta}_i,{\vec\gamma}_{i,j})d\rho_{c,i} d{\vec\beta}_i d{\vec\gamma}_{i,j}\; .
\label{eq:posterior}
\end{equation}
The proportionality in this expression is to remind us that we will need, as a final step, to normalize $P({\vec\alpha})$ so that $\int P({\vec\alpha})d{\vec\alpha}=1$.  The likelihood ${\cal L}$ is the product of all of the individual likelihoods, so
\begin{equation}
{\cal L}({\vec\alpha},\rho_{c,i},{\vec\beta}_i,{\vec\gamma}_{i,j})=\prod_{i,j}{\cal L}_{i,j}({\vec\alpha},\rho_{c,i},{\vec\beta}_i,{\vec\gamma}_{i,j})\; ,
\end{equation}
where ${\cal L}_{i,j}({\vec\alpha},\rho_{c,i},{\vec\beta}_i,{\vec\gamma}_{i,j})$ is the likelihood of measuring data set $j$ from star $i$ given the model under consideration with parameter values ${\vec\alpha}$, $\rho_{c,i}$, ${\vec\beta}_i$, and ${\vec\gamma}_{i,j}$.

We make the following two simplifying assumptions:

Assumption 1: the prior $q({\vec\alpha},\rho_{c,i},{\vec\beta}_i,{\vec\gamma}_{i,j})$ in expression (\ref{eq:posterior}) can be represented as the product of the following factors:
\begin{equation}
q({\vec\alpha},\rho_{c,i},{\vec\beta}_i,{\vec\gamma}_{i,j})=q({\vec\alpha})\left[\prod_i q(\rho_{c,i}|{\vec\alpha})\right]\left[\prod_i q({\vec\beta}_i)\right]\left[\prod_{i,j}q({\vec\gamma}_{i,j})\right]\; .
\end{equation}
Thus, we assume that the priors are independent of each other, with the exception of the prior on the central density.  We write the prior on $\rho_{c,i}$ as $q(\rho_{c,i}|{\vec\alpha})$ because it is possible that the prior will depend on other parameters (for example, for a nonrotating star, the maximum central density of a stable star will often depend only on ${\vec\alpha}$, but in general, the maximum stable density will also depend on the rotation rate).  In principle, other parameters could also be codependent, e.g., if one of our parameters is the rotation frequency, then its maximum value depends on both $\rho_{c,i}$ and ${\vec\alpha}$.  However, for the cases we consider here, the rotation frequency is small enough that it is unimportant.

Assumption 2: We assume that when we break the overall likelihood into a product of the likelihoods of the individual data sets given the model and parameter values, the parameters not associated with a given observed quantity do not affect the likelihood of the measured value of that quantity.  For example, for a given distribution of central densities, we assume that the central density of one star has no influence on the likelihoods of the measured values of the parameters that describe another star.  This means we can write
\begin{equation}
\begin{array}{rl}
{\cal L}({\vec\alpha},\rho_{c,i},{\vec\beta}_i,{\vec\gamma}_{i,j})&=\prod_{i,j}{\cal L}_{i,j}({\vec\alpha},\rho_{c,i},{\vec\beta}_i,{\vec\gamma}_{i,j})\\
&=\prod_i \left[\prod_{j|i}{\cal L}_{i,j}({\vec\alpha},\rho_{c,i},{\vec\beta}_i,{\vec\gamma}_{i,j})\right]\; .\\
\end{array}
\end{equation}
Here, ``$j|i$" means ``the set $j$ of measurements of star $i$".

These assumptions allow us to write the posterior probability density for ${\vec\alpha}$ as follows:
\begin{equation}
P({\vec\alpha})\propto q({\vec\alpha})\prod_i\left[\int q(\rho_{c,i}|{\vec\alpha})q({\vec\beta}_i)\prod_{j|i}\left(\int q({\vec\gamma}_{i,j}){\cal L}_{i,j}({\vec\alpha},\rho_{c,i},{\vec\beta}_i,{\vec\gamma}_{i,j})d{\vec\gamma}_{i,j}\right)d\rho_{c,i} d{\vec\beta}_i\right]\; .
\label{eq:margposterior}
\end{equation}
When we compare this with the general expression $P({\vec\alpha})\propto q({\vec\alpha}){\cal L}({\vec\alpha})$ we see that, given our assumptions, the likelihood of the full set of all data given the model and parameter values ${\vec\alpha}$ is
\begin{equation}
{\cal L}({\vec\alpha})=\prod_i\left[\int q(\rho_{c,i}|{\vec\alpha})q({\vec\beta}_i)\prod_{j|i}\left(\int q({\vec\gamma}_{i,j}){\cal L}_{i,j}({\vec\alpha},\rho_{c,i},{\vec\beta}_i,{\vec\gamma}_{i,j})d{\vec\gamma}_{i,j}\right)d\rho_{c,i} d{\vec\beta}_i\right]\; .
\label{eq:likelihood}
\end{equation}

Loosely speaking, this approach assigns the likelihood of each set of values of the measured quantities, given the data, to all combinations of the model parameter values that yield these values of the measured quantities.  To see why this is appropriate, note that when we analyze particular neutron-star data, we find that the central density and EOS parameters only influence a subset of the parameters that are used to describe the data.  For example, the distance, direction, and orientation of a merging binary do not depend on either $\rho_{c,i}$ or ${\vec\alpha}$.  Similarly, when energy-dependent X-ray waveforms from {\it NICER} are analyzed, only the gravitational mass $M$ and circumferential radius $R$ depend on $\rho_{c,i}$ and ${\vec\alpha}$.  Thus, in, e.g., the waveform case, the marginalized likelihood associated with given $\rho_{c,i}$ and ${\vec\alpha}$ will be the same as the corresponding marginalized likelihood associated with the corresponding $M$ and $R$, where the marginalization is performed over all of the other parameters that describe the particular data set.

Once we have the posterior density at each of a large number of EOS parameter combinations, we compute the posterior density in pressure at a specific density $\rho_0$ by (1)~determining the pressures predicted at $\rho_0$ using each parameter combination, (2)~assigning a statistical weight to each pressure that is the same as the posterior density for the parameter combination, and then (3)~sorting the predicted pressures at $\rho_0$ in increasing order.  We then determine a given credibility quantile (e.g., the 5\% quantile of the pressure at $\rho_0$) by summing the normalized weights of the pressures at $\rho_0$ until 5\% is reached.  

\section{Using Different Types of Measurements}
\label{sec:specific}

Different measurements and observations require different approaches to use them in our statistical procedure for constraining the EOS.  Some, such as the nuclear symmetry energy, can be computed directly from the EOS for a broad category of nuclear models.  Others, such as the binary tidal deformability measured for GW170817 or future events, require marginalization.  We now discuss illustrative measurements and observations and how they can be used in our statistical procedure.  As we discussed in Section~\ref{sec:statistics}, we can obtain the likelihood from multiple independent measurements by simply multiplying their individual likelihoods.  We also note that additional measurements beyond what we consider here, such as measurements of neutron-star quasinormal modes (e.g., Kokkotas  \& Schmidt 1999), may be available in the future.

\subsection{Constraints Not Requiring Marginalization}

\subsubsection{Nuclear Symmetry Energy}  

In our discussion here, we assume that the nuclear symmetry energy $S$ is the difference in the energy per nucleon between pure neutron matter (which at density $n$ we denote by $\epsilon(n)/n$) and symmetric nuclear matter, at the nuclear saturation density $n_s$.  We are interested in the EOS of cold catalyzed matter, which is not purely neutrons. However, as pointed out by \citet{2016PhR...621..127L} the proton fraction at $n_s$ is only $\sim 1$\%, which is small enough to be neglected.  With this approximation $S=\epsilon(n_s)/n_s-E_{\rm sym}$, where $E_{\rm sym}=-16.0$~MeV is the energy per nucleon of symmetric matter at nuclear saturation density \citep{2012PhRvC..86a5803T}.  If the measured value of $S$ is $S_0$ and the predicted value for EOS parameters ${\vec\alpha}$ is $S({\vec\alpha})$, then the likelihood factor associated with the symmetry energy is simply
\begin{equation}
{\cal L}_S({\vec\alpha})={\cal L}(S_0|S({\vec\alpha}))\; .
\end{equation}

\subsubsection {Gravitational Mass}

We can in principle obtain information about the EOS from any measurement of a neutron-star mass.  For example, if an EOS has a maximum mass of $2.5~M_\odot$ but no neutron stars are found to have masses larger than $2.2~M_\odot$, that EOS could be disfavored (we thank R. Essick for emphasizing this point to us).  However, the mass distribution of neutron stars depends on more than the EOS.  For instance, although all equations of state allow $0.5~M_\odot$ neutron stars to exist, there are no plausible suggestions for how such stars can form.  Moreover, the path to forming high-mass neutron stars is not well understood; in the example above, it could be that it is simply extremely rare that a star's birth and subsequent accretion will produce a mass above $2.2~M_\odot$ even if significantly higher masses are allowed by the EOS.  A complete analysis would take all measured masses into account using a joint, parameterized model of birth and accretion as well as the EOS, but this is not currently feasible.  We therefore focus on the highest measured masses.

In the limit of slow rotation, the maximum gravitational mass is a function only of the EOS.  It is the gravitational mass $M_{\rm max}$ at the largest central total mass$-$energy density $\rho_c$ such that $dM/d\rho_c\geq 0$.  

If the posterior probability distribution for the mass of star $j$ is $P(M_j)$, then the likelihood factor for the EOS parameter values ${\vec\alpha}$ for that star is
\begin{equation}
{\cal L}_{M_j}({\vec\alpha})=\int_0^{M_{\rm max}({\vec\alpha})}P(M_j)dM\; .
\end{equation}
A similar integral can take into account observations that disfavor {\it large} maximum masses, and in Section~\ref{sec:results} we show the results for one such hypothetical constraint.

\subsubsection{Moment of Inertia}  

For a given EOS, the expected moment of inertia can be computed given either a central density or a mass \citep{1967ApJ...150.1005H}.  If we assume that we know the mass $M_0$ very precisely (as is the case for both components of the double pulsar PSR~J0737$-$3039, which is the system of greatest promise for moment-of-inertia measurements), then when a measurement is made of the moment of inertia of the pulsar, the likelihood factor will be
\begin{equation}
{\cal L}_I({\vec\alpha})={\cal L}(I_{\rm obs}|I({\vec\alpha},M=M_0))\; ,
\end{equation}
where ${\cal L}(I_{\rm obs}|I({\vec\alpha},M=M_0))$ is the likelihood of observing a moment of inertia $I_{\rm obs}$ if the expected value at $M=M_0$ is $I({\vec\alpha})$ for EOS parameter values ${\vec\alpha}$.

\subsubsection{Gravitational Binding Energy}  

Suppose that a star with a precisely measured gravitational mass $M_0$ is thought to have a baryonic mass $M_{\rm bary,0}$ (and thus a binding energy $M_{\rm bary,0}c^2-M_0c^2$) with some likelihood ${\cal L}(M_{\rm bary,0}|M_0)$ (one such possible scenario is if there is evidence that the neutron star was formed in an electron-capture supernova (\citealt{1984ApJ...277..791N,2004ApJ...612.1044P,2005MNRAS.361.1243P,2019arXiv190704184Z}; see Section~\ref{sec:results} for details and caveats).  Then
\begin{equation}
{\cal L}_{E_{\rm bind}}({\vec\alpha})={\cal L}(M_{\rm bary,0}|M_{\rm bary}({\vec\alpha},M_0)) ,
\end{equation}
where $M_{\rm bary}({\vec\alpha},M_0)$ is the baryonic mass for a gravitational mass $M_0$ that is predicted using the EOS with parameter values ${\vec\alpha}$.

\subsection{Constraints Requiring Marginalization}

\subsubsection{Binary Tidal Deformability in Neutron-star Mergers}

The newest category of EOS-relevant neutron-star observations is the constraint on the tidal deformability of neutron stars that has been obtained using gravitational-wave observations of GW170817 \citep{2019PhRvX...9a1001A}.  The dimensionless form of the tidal deformability, for a star of gravitational mass $M$ and circumferential radius $R$, is
\begin{equation}
\Lambda={2\over 3}k_2\left(Rc^2\over{GM}\right)^5\; .
\end{equation}
Here, $k_2$ is the tidal Love number.  \citet{2008ApJ...677.1216H} has a good discussion of how to compute $\Lambda$ given an EOS and the central density (see also the erratum at \citealt{2009ApJ...697..964H}).  Gravitational-wave measurements give a tighter constraint on the binary tidal deformability than on the tidal deformabilities of the two stars individually: indeed, at least for the Taylor family of post-Newtonian waveforms, the most easily measurable quantity for stars of masses $M_1$ and $M_2\leq M_1$ with tidal deformabilities $\Lambda_1$ and $\Lambda_2$ is \citep{2014PhRvD..89j3012W}
\begin{equation}
{\tilde\Lambda}={16\over{13}}{(M_1+12M_2)M_1^4\Lambda_1+(M_2+12M_1)M_2^4\Lambda_2\over{(M_1+M_2)^5}}\; .
\end{equation}
In such events, the masses are not measured well individually, but the chirp mass $M_{\rm ch}=(M_1M_2)^{3/5}/(M_1+M_2)^{1/5}$ is known precisely; for example, for GW170817, $M_{\rm ch}=1.186\pm 0.001~M_\odot$ \citep{2019PhRvX...9a1001A}.  We note that, for fixed $M_{\rm ch}$, ${\tilde\Lambda}$ is relatively insensitive to the mass ratio $M_2/M_1$.  For instance, using the scaling $\Lambda\propto M^{-6}$ suggested by \citet{2018PhRvL.121i1102D}, ${\tilde\Lambda}$ for $M_2/M_1=0.6$ is only $\sim 5\%$ larger than ${\tilde\Lambda}$ for $M_2/M_1=1$.

Because only $M_{\rm ch}$ is measured precisely, we need to marginalize over the masses.  We approach this marginalization problem by assuming that gravitational-wave data analysis has given us a full posterior in $(M_1,M_2,{\tilde\Lambda})$ space.  For given EOS parameter values, the prior probability distribution for the masses is set by the prior probability distribution for the central densities (or by equivalent criteria).  For fixed EOS parameter values, we can compute ${\tilde\Lambda}={\tilde\Lambda}(M_1,M_2,{\vec\alpha})$.  Thus, in general, we would compute this likelihood factor by integrating over both $M_1$ and $M_2$:
\begin{equation}
{\cal L}_\Lambda({\vec\alpha})=\int dM_1\int q(M_1)q(M_2){\cal L}(M_1,M_2,{\tilde\Lambda}|{\vec\alpha})dM_2\; ,
\end{equation}
where $q(M_1)$ and $q(M_2)$ are the priors for $M_1$ and $M_2$ and ${\cal L}(M_1,M_2,{\tilde\Lambda}|{\vec\alpha})$ is the three-dimensional likelihood obtained from the analysis of the gravitational-wave data, given EOS parameter values ${\vec\alpha}$.

However, $M_{\rm ch}$ is known with such high precision and accuracy that, given a value for $M_1$, $M_2$ is known to high accuracy.  Therefore, we can recast the likelihood factor as
\begin{equation}
{\cal L}_\Lambda({\vec\alpha})=\int dM_1 q(M_1)\int q(M_2|M_{\rm ch},M_1){\cal L}(M_1,M_2,{\tilde\Lambda}|{\vec\alpha})dM_{\rm ch}\; ,
\end{equation}
where $q(M_2|M_{\rm ch},M_1)$ is the prior probability density for $M_2$ at the value of $M_2$ implied by $M_{\rm ch}$ and $M_1$, and the integral is over the probability distribution for $M_{\rm ch}$ obtained from the gravitational-wave analysis.  Note that even if $M_{\rm ch}$ is known with high precision, we cannot write the second integral as a delta function.  This is because the total probability in the narrow range of $M_2$ allowed for a given $M_{\rm ch}$ and $M_1$ depends on the EOS and the prior for the masses (or central densities).  As a result, this factor must be calculated directly for each EOS.

\subsubsection{Radius and Mass}

Suppose that for a given star the likelihood of a mass $M$ and radius $R$ is ${\cal L}(M,R)$.   For a given stellar mass, the radius $R$ is determined precisely for given EOS parameter values.  Thus, the likelihood factor associated with a radius measurement is 
\begin{equation}
{\cal L}_{R}({\vec\alpha})=\int dM q(M){\cal L}_l(M,R(M,{\vec\alpha}))\; ,
\end{equation}
where $R(M,{\vec\alpha})$ is the circumferential radius for a gravitational mass $M$ given EOS parameter values ${\vec\alpha}$, and $q(M)$ is the prior on $M$.  Note that the integration is equivalent to integrating the full $(M,R)$ likelihood over the full $(M,R)$ curve predicted using a given EOS.  

\subsection{Combination of Constraints}

Under the assumption of independent measurements that we described earlier, we can determine the final likelihood ${\cal L}({\vec\alpha})$ at a given set of values of the EOS parameters ${\vec\alpha}$ by simply setting it equal to the product of the individual likelihoods.  Thus, if there is some set $i$ of independent symmetry energy measurements, some set $j$ of neutron-star mass measurements high enough to be constraining (noting that here the use of $i$ and $j$ is different than it was in Section~\ref{sec:statistics}), some set $k$ of binary tidal deformability measurements, some set $l$ of mass$-$radius pairs, some set $m$ of moments of inertia, and some set $n$ of gravitational binding energies, then the final likelihood is
\begin{equation}
{\cal L}({\vec\alpha})=\left[\prod_i {\cal L}_{S,i}({\vec\alpha})\right]\left[\prod_j {\cal L}_{M_j}({\vec\alpha})\right]\left[\prod_k {\cal L}_{\Lambda,k}({\vec\alpha})\right]\left[\prod_l {\cal L}_{R,l}({\vec\alpha})\right]\left[\prod_m {\cal L}_{I,m}({\vec\alpha})\right]\left[\prod_n {\cal L}_{E_{\rm bind},n}({\vec\alpha})\right]\; .
\end{equation}
We stress that this expression implicitly assumes that systematic errors can be neglected.  If they cannot, then --- as always --- there is the prospect for significant bias.

\section{Comparison with Previous Approaches}
\label{sec:previous}

In this section, we compare our statistical method with EOS constraint methods in the literature.  In Section~\ref{sec:results} we will discuss specific inferences of masses, radii, etc. that are then used to constrain the EOS.  Here, we focus on the statistical approaches themselves.  Our method is generally consistent with other methods that are fully Bayesian, e.g., among recent papers \citet{2015PhRvD..91d3002L}, \citet{2015PhRvD..92b3012A}, \citet{2016EPJA...52...69A}, and \citet{2018MNRAS.478.1093R}.  The non-parameteric approach of \citet{2019PhRvD..99h4049L} is also worth consideration.

\subsection{Use of Bounds in Mass or Other Quantities}

As we have emphasized, in a fully consistent Bayesian analysis, a given observation needs to be incorporated using a likelihood-based procedure.  Imposing a strict bound of any kind, other than bounds stemming from fundamental physical laws, may discard important information. However, to our knowledge, all previous analyses except that of \citet{2016EPJA...52...69A} have used a hard lower bound on the maximum mass, in the sense that a given EOS or parameter combination is allowed if it has a maximum mass above some specified value (often $1.97~M_\odot$, because the $M=2.01\pm 0.04~M_\odot$ mass reported by \citealt{2013Sci...340..448A} for PSR~J0347+0432 was the highest reported mass until the $M=2.14^{+0.10}_{-0.09}~M_\odot$ mass reported by \citealt{2019arXiv190406759T} for PSR~J0740+6620), and disallowed if the maximum mass is below the bound.  A similar approach is taken commonly, but not as universally, with the tidal deformability measurement from GW170817.  

The first reason that this is incorrect is illustrated nicely by the progression in time of the estimates of the mass of PSR~1614$-$2230.  The first measurement, by \citet{2010Natur.467.1081D}, was $M=1.97\pm 0.04~M_\odot$.  The second measurement, by \citet{2016ApJ...832..167F}, was $M=1.928\pm 0.017~M_\odot$.  The most recent measurement, by \citet{2018ApJ...859...47A}, is $M=1.908\pm 0.016~M_\odot$.  Thus, the best estimate of the mass in both updates is slightly more than one standard deviation lower than the previous best estimate.  Thus, a strict lower bound at the $-1\sigma$ mass $M=1.93~M_\odot$ from the first measurement would be too restrictive given our current knowledge of the mass of PSR~1614$-$2230.  Instead, one should use the full posterior distribution of the mass.

The second reason why this approach is suboptimal is that there is, after all, uncertainty in the mass measurements.  If we accept $M=2.14^{+0.10}_{-0.09}~M_\odot$ as the mass estimate for PSR~J0740+6620, then using the hard-bound approach, an EOS with a maximum mass of $2.05~M_\odot$ is just as viable as an EOS with a maximum mass of $2.14~M_\odot$.  But if we assume that the measurement has only Gaussian statistical uncertainties, there is an $\sim$84\% probability that the mass of PSR~J0740+6620 is {\it greater} than $2.05~M_\odot$.  Thus, in reality, the EOS with $M_{\rm max}=2.14~M_\odot$ is considerably more consistent with the data than the EOS with $M_{\rm max}=2.05~M_\odot$.  Applying a lower bound is not a statistically appropriate approach.

The third reason that strict bounds should not be used is that this approach does not allow the incorporation of information from multiple stars.  For example, at the moment, the only published masses that pose significant constraints to the EOS are $M=2.14^{+0.10}_{-0.09}~M_\odot$ for PSR~J0740+6620, $M=2.01\pm 0.04~M_\odot$ for PSR~J0348+0432, and $M=1.908\pm 0.016~M_\odot$ for PSR~1614$-$2230.  An EOS with $M_{\rm max}=1.8~M_\odot$ is disfavored at the $3.8\sigma$ level for PSR~J0740+6620 alone, but at more than $9\sigma$ when measurements of the masses of all three pulsars are included (using the simple assumption that the uncertainties are exactly Gaussian, which is unlikely to be true at several standard deviations).  Thus, $M_{\rm max}=1.8~M_\odot$ is excluded much more strongly based on the data from all three stars than it would be using just the most massive of the three.  If a future star is discovered with, say, a mass $M=2.01\pm 0.05~M_\odot$, then using the hard-bound method, it would not contribute at all to EOS constraints, whereas in reality, it would make low-$M_{\rm max}$ EOSs significantly less probable.

\subsection{Lack of Marginalization}

It is common, although not universal, for post-GW170817 EOS constraint papers to use an estimate of the tidal deformability parameter at $1.4~M_\odot$ in constraints, rather than integrating over the full posterior space.  Similarly, numerous papers use only the maximum-likelihood or minimum-$\chi^2$ point along the $R(M)$ curve implied by a given EOS, whereas the integration should instead be performed over the whole curve (see, for example, \citealt{2010ApJ...722...33S,2016ApJ...820...28O}).

\subsection{Attempts to Invert Measurements to Obtain the EOS}

Early papers on inference of the EOS from neutron-star measurements often presented EOS determination as an inversion of neutron-star measurements, sometimes using a Jacobian formalism to map neutron-star observables into EOS parameters.  Such an approach misses the fact that this is intrinsically a measurement problem, not a problem of inverting a mathematical relation, and thus must be approached statistically.  Not approaching the analysis as a measurement problem can lead to fundamental difficulties.

Even setting aside for the moment the fundamentally statistical nature of the problem, in realistic situations, attempts to invert observed quantities to determine EOS parameter values fail because the inversion is singular.  For example, if two $M(R)$ curves obtained from different equations of state cross, then the inversion is clearly singular at the crossing point.  Another difficulty with this approach has been emphasized by \citet{2018MNRAS.478.1093R} and \citet{2018MNRAS.478.2177R}, in the context of EOS models that have separately parameterized segments at different densities, such as models that use a sequence of polytropes.  They point out that some neutron stars might not have a central density large enough to reach the highest density in the EOS model.  In that case, the parameters describing higher densities have no influence on the mass and radius of that star, and thus nothing can be inferred about those parameters \citep{2018MNRAS.478.2177R}.  A further difficulty with approaching EOS parameter estimation as a mathematical inversion problem rather than as statistical inference is that a one-to-one mapping requires that the number of EOS parameters be equal to the number of observables.  Of course, the hope is that there are many more observations than model parameters!  

For these reasons, most papers in the last decade have approached this problem correctly, as a statistical inference problem, rather than as a problem of inverting a
map between observations and model parameters.

\section{Results}
\label{sec:results}

In this section, we present the 5\%, 50\%, and 95\% credibility quantiles for the pressure at a set of densities and for the circumferential radii at a set of gravitational masses, obtained using progressively more restrictive data.  The densities start at half of nuclear saturation density (i.e., at 0.08 baryons per fm$^3$), where the pressures of all of our EOS models agree by construction, because up to that density we use the SLy \citep{2001A&A...380..151D} EOS.  We then construct the cumulative probability distribution for the pressure at progressively higher densities.  We also plot the $M-R$ curves that bound the region that makes up 90\% of the total probability.  Currently, constraints on the EOS are relatively weak, which means that most EOS parameter combinations have high likelihoods, and thus, we do not need to perform sophisticated searches through parameter space. 

Our method can be used with any parameterization of the EOS.  We assume that the pressure is a function only of the density, i.e., that the EOS is barotropic.  The pressure does not depend explicitly on the temperature or the proton fraction, because we assume that the matter is in beta equilibrium. For our primary parameterization we follow \citet{2018PhRvL.121p1101A} in using the spectral parameterization introduced by Lindblom \citep{2010PhRvD..82j3011L,2018PhRvD..97l3019L}, in which the free parameters are spectral indices $\gamma_k$ that represent the adiabatic index $\Gamma(p)=[(\rho+p)/p](dp/d\rho)$ (where $p$ is the pressure and $\rho$ is the total mass$-$energy density) using the expansion
\begin{equation}
\Gamma(p)=\exp\left(\sum_k\gamma_kx^k\right)\; ,
\end{equation}
where $x\equiv\log(p/p_0)$ and $p_0$ is the pressure at half of nuclear saturation density.  We also follow previous work (e.g., \citealt{2018PhRvL.121p1101A,2018PhRvD..98f3004C}) by using an expansion up to $x^3$ with the following uniform priors on the coefficients $\gamma_k$: $\gamma_0\in[0.2,2]$, $\gamma_1\in[-1.6,1.7]$, $\gamma_2\in[-0.6,0.6]$, and $\gamma_3\in[-0.02,0.02]$.  We do not additionally require, as some papers have, that $\Gamma(p)\in[0.6,4.5]$ at all densities.  The parameterization itself guarantees that $\Gamma(p)>0$, which is needed to enforce thermodynamical stability.  We also require that the adiabatic speed of sound be less than the speed of light.  In Section~\ref{sec:polytrope} we display results using an alternative parameterization which has potentially different polytropic indices at variable transition densities.

Once the EOS is chosen, then in the slow rotation limit, the mass and radius as functions of the central density, the maximum stable mass, and the gravitational binding energy for a given gravitational mass follow from the Tolman-Oppenheimer-Volkoff (TOV) equation \citep{1939PhRv...55..364T,1939PhRv...55..374O}, and from the relation between baryonic mass density and total mass$-$energy density discussed in \citet{1965ApJ...142.1541T}.  We compute the moment of inertia and spin quadrupole moment following the approach in \citet{1967ApJ...150.1005H}, and the tidal Love number using the development in \citet{2008ApJ...677.1216H} (see also the erratum at \citealt{2009ApJ...697..964H}).  We verified the accuracy of our code by comparing our outputs with those listed in Table~III of \citet{2009PhRvD..79l4032R} (using their equation of state rather than the spectral parameterization).  We also checked that our moments of inertia, quadrupole moments, and tidal deformabilities follow closely the I-Love-Q relations (\citealt{2013PhRvD..88b3009Y} and subsequent papers).

The order in which we add measurements is (1)~symmetry energy (from laboratory measurements), (2)~mass measurements, (3)~tidal deformability measurements, (4)~hypothetical future measurements of both radius and mass, (5)~hypothetical future measurements of moments of inertia, and finally (6)~hypothetical future measurements of the binding energy of stars with precisely measured gravitational masses.  That is, in the first section we present results assuming only measurements of (1) (with different illustrative levels of precision for the symmetry energy).  We then present results assuming only measurements of (1) and (2) (with a standard precision for the symmetry energy and different potential measurements for the mass), and so on.  This makes it possible to see how additional measurements progressively improve the precision of our understanding of the EOS and, as a consequence, the neutron-star mass$-$radius relation.  Note that when a new measurement is incorporated, the new EOS constraints can shift beyond the previous 5\% or 95\% quantile.  For example, if a neutron star is measured to have a high mass then soft equations of state are disfavored, which then shifts the quantiles to higher pressure at a given density.  

Whereas in Section~\ref{sec:statistics} we presented our general statistical method, and in Section~\ref{sec:specific} we discussed how to apply our method to particular types of measurements, here we use both existing and potential future measurements to find credibility regions in $P(\rho)$ space.  Thus we need to make choices about which measurements to use.  For example, after GW170817, there have been many detailed simulations and comparisons with electromagnetic information (especially the details of the resulting kilonova) that have endeavored to constrain the maximum mass of neutron stars, or to place lower limits on the tidal deformability of neutron stars of particular masses.  We also need to specify the prior on the mass or the central density for a given combination of EOS parameter values.  In the results we present here, we assume that the central density can with equal probability be anywhere between the density that would produce an $M=1.0~M_\odot$ neutron star with that EOS, and the density that produces the maximum mass possible for that EOS.  We stress that although we make particular choices, these are only illustrative.  Our focus is not to produce our own version of the constraints, although given our assumptions, our current constraints are in the top left panels of Figure~\ref{fig:EOSSMmaxL} and Figure~\ref{fig:RMSMmaxL}.  Instead, we make these choices to demonstrate how our method works in practice; other choices of measurements and even of EOS families and the priors on their parameters could be used straightforwardly with our method.

Our final note prior to presenting our results is a reminder that all measurements and observations have to be interpreted within a model framework, and this means that we rely on that framework to obtain quantities of interest.  For example, virtually all neutron-star observations are interpreted under the assumption that general relativity properly describes extreme gravity.  Many papers prior to the direct detection of gravitational waves pointed out that the mass$-$radius relation (and thus all other structural aspects of stars) could be modified considerably in different theories of gravity (see \citealt{2003PhRvL..90n1101D} and \citealt{2013GReGr..45..771O} for just two examples).  Careful analysis of gravitational-wave data has limited the prospects for deviations from general relativity in stellar-mass objects (see \citealt{2016PhRvD..94h4002Y} for an excellent summary after the first two events), but it is useful to keep an open mind. 

\begin{figure}
\begin{center}
  \resizebox{1.0\textwidth}{!}{\includegraphics{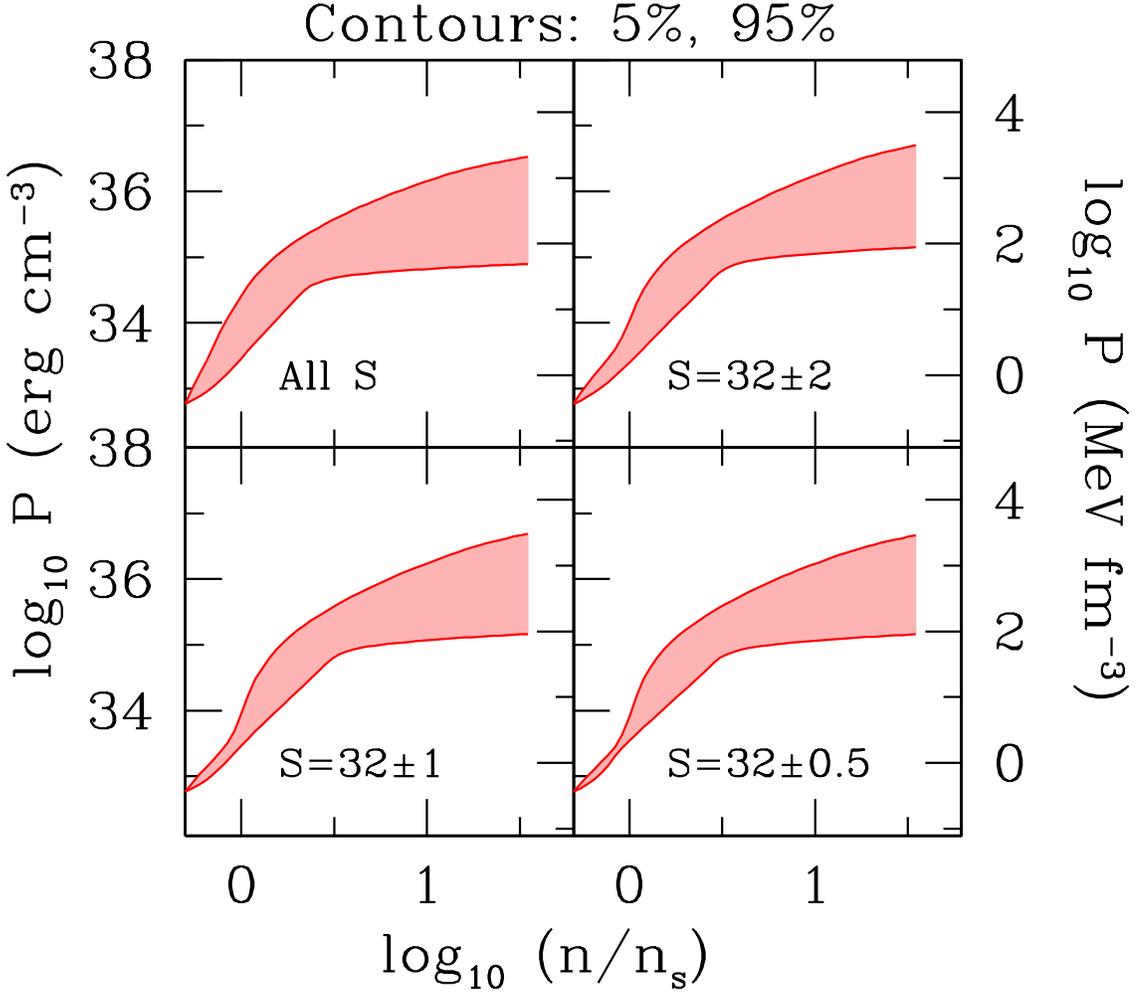}}
\vspace{-2.5truein}
   \caption{Equation-of-state constraints based only on symmetry energy measurements.  In this figure, as in the $P-\rho$ figures that follow, the bottom red curve shows the 5\% quantile in the pressure posterior at each density and the top red curve shows the 95\% quantile.  The shaded region is between the 5\% and 95\% quantiles.  All figures also give the $\log_{10}$ of the effective number density (which we define as the rest-mass density divided by the mass of a neutron) in units of the number density at nuclear saturation ($n_s\approx 0.16$~fm$^{-3}$) and the $\log_{10}$ of the pressure in erg~cm$^{-3}$ on the left-hand axes and in MeV~fm$^{-3}$ on the right-hand axes.  The top left panel shows the constraints when all values of the symmetry energy $S$ are considered equally probable.  It therefore shows the 5\%$-$95\% range of the prior.  The top right panel applies a Gaussian likelihood to the symmetry energy, with mean $S_{\rm mean}=32$~MeV and standard deviation $\sigma_S=2$~MeV; the bottom left panel uses a Gaussian likelihood with $S_{\rm mean}=32$~MeV and $\sigma_S=1$~MeV; and the bottom right panel uses a Gaussian likelihood with $S_{\rm mean}=32$~MeV and $\sigma_S=0.5$~MeV.  As expected, more precise symmetry energy measurements tighten the EOS at $n_s$ and below, but have little impact on the EOS at higher densities.}
\label{fig:EOSS}
\end{center}
\end{figure}

\begin{figure}
\begin{center}
  \resizebox{1.0\textwidth}{!}{\includegraphics{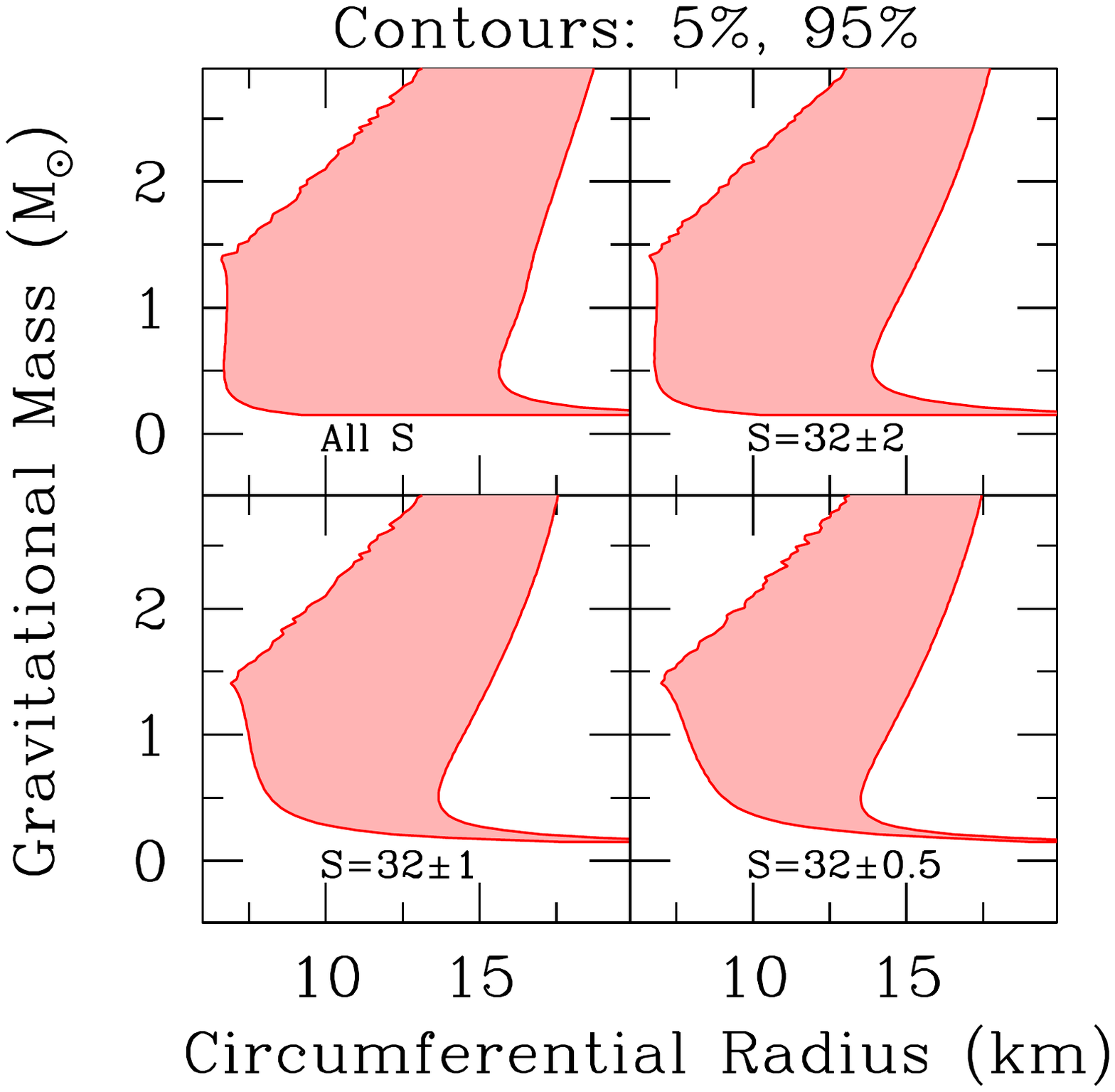}}
\vspace{-2.5truein}
   \caption{Mass$-$radius curves following from the equation-of-state constraints shown in Figure~1.  In this figure, as in the $M-R$ figures that follow, the left red curve and right red curve are the lower and upper boundaries, respectively, of the envelope of equation-of-state curves that make up 90\% of the total probability in the sample.  The central densities for even the $1.0~M_\odot$ stars are well above nuclear saturation density, so constraints on $S$ have little impact on the radius of stars with realistic masses.}
\label{fig:RMS}
\end{center}
\end{figure}

\subsection{Nuclear Symmetry Energy}

\citet{2012PhRvC..86a5803T} give the status of a number of different laboratory measurements that could constrain the nuclear symmetry energy.  We treat the likelihood factor from the symmetry energy as a Gaussian:
\begin{equation}
{\cal L}_S({\vec\alpha})={1\over{(2\pi\sigma_S^2)^{1/2}}}e^{-(S({\vec\alpha})-S_0)^2/2\sigma_S^2}\; ,
\end{equation}
where $S({\vec\alpha})$ is the symmetry energy predicted using specified values of the EOS parameters ${\vec\alpha}$.  For our standard constraint, we choose $S_0=32$~MeV and $\sigma_S=2$~MeV from a rough averaging of the various results presented in \citet{2012PhRvC..86a5803T}.  Non-Gaussian likelihoods are also straightforward to include in our framework.  

Figure~\ref{fig:EOSS} shows that more precise measurements of $S$ would strongly constrain the EOS below nuclear saturation density but would have little effect above $n_s$.  Figure~\ref{fig:RMS} shows that knowledge of $S$ has little impact on our knowledge of the radius of stars with $M>1.0~M_\odot$.

\begin{figure}
\begin{center}
  \resizebox{1.0\textwidth}{!}{\includegraphics{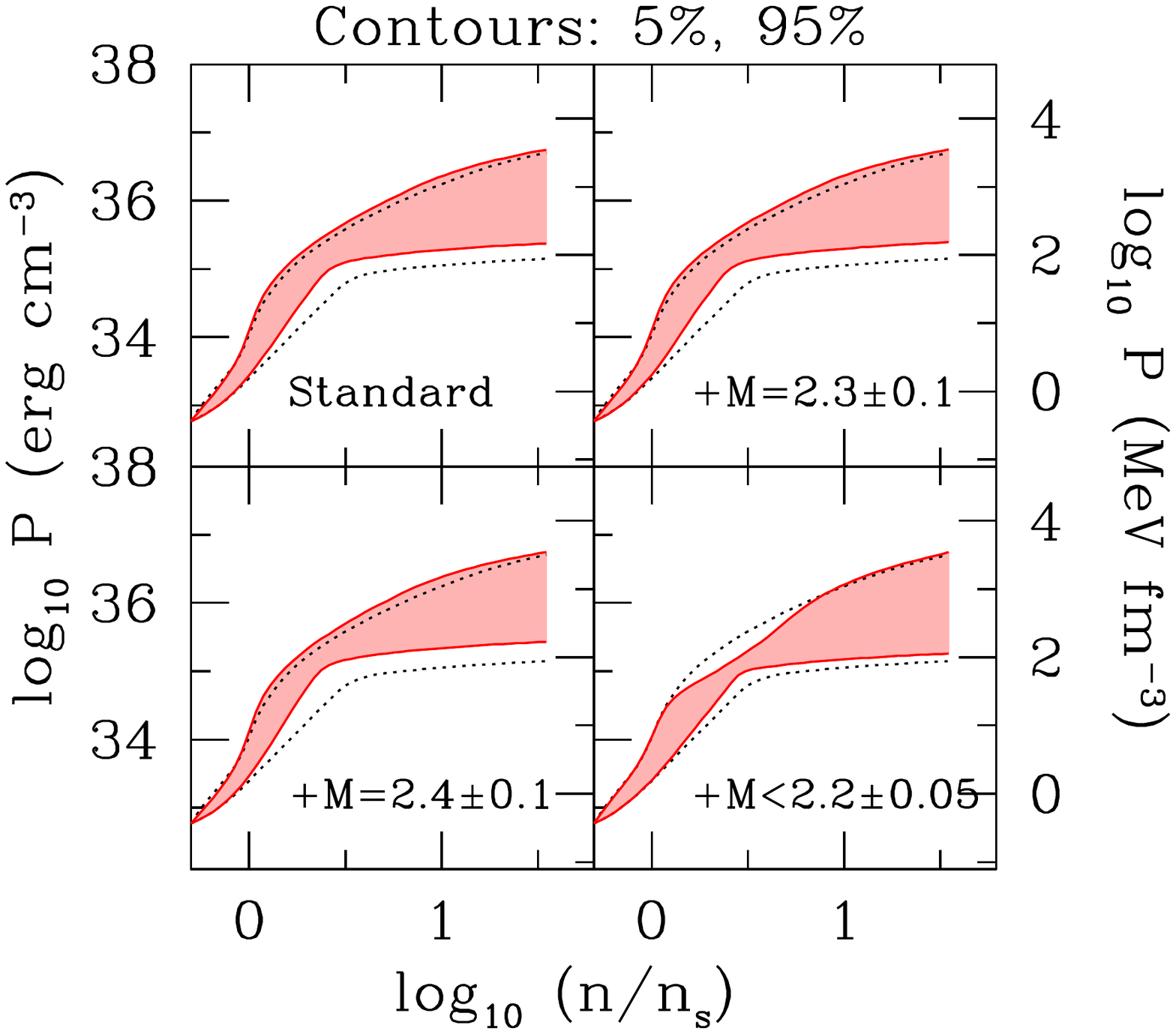}}
\vspace{-2.5truein}
   \caption{Equation-of-state constraints based on symmetry energy and mass measurements.  Here, we assume that the probability distribution for the symmetry energy is a Gaussian with mean 32~MeV and standard deviation 2~MeV, and the dotted lines show the 5\% and 95\% quantiles at each density when only the symmetry energy is used as a constraint (with $S=32\pm 2$~MeV).  The top left panel shows the quantiles when we include constraints based on the masses of PSR~J0740+6620 ($M=2.14^{+0.10}_{-0.09}~M_\odot$; see \citealt{2019arXiv190406759T}), PSR~J0348+0432 ($M=2.01\pm 0.04~M_\odot$; see \citealt{2013Sci...340..448A}), and PSR~J1614$-$2230 ($M=1.908\pm 0.016~M_\odot$; see \citealt{2016ApJ...832..167F}).  The top right panel shows the effect of adding, to those three stars, a hypothetical star with a mass measurement of $M=2.3\pm 0.1~M_\odot$.  The bottom left panel shows the effect of adding instead a star with $M=2.4\pm 0.1~M_\odot$.  The bottom right panel shows the effect of adding instead an {\it upper} limit of $M=2.2\pm 0.05~M_\odot$ to the maximum mass, from arguments about short gamma-ray bursts and events such as GW170817 \citep{2013PhRvL.111m1101B,2015ApJ...812...24F,2015ApJ...808..186L,2017ApJ...850L..19M}.  Mass measurements significantly constrain the EOS below $\sim 10n_s$.}
\label{fig:EOSSMmax}
\end{center}
\end{figure}

\begin{figure}
\begin{center}
  \resizebox{1.0\textwidth}{!}{\includegraphics{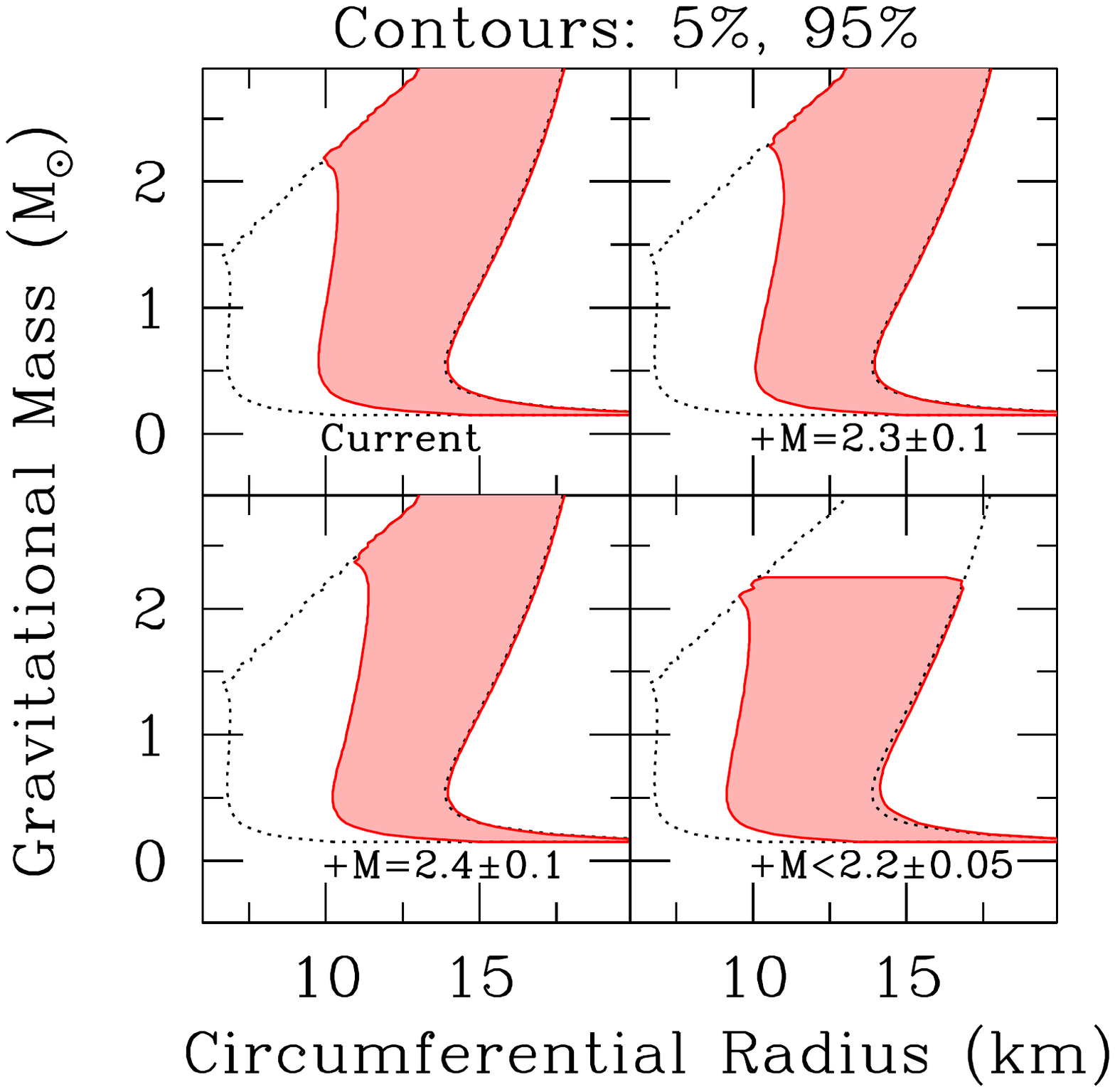}}
\vspace{-2.5truein}
   \caption{Mass$-$radius constraints based on symmetry energy and mass measurements.  The constraints shown in each panels correspond to the EOS constraints shown in the corresponding panel of Figure~\ref{fig:EOSSMmax}, and the dotted lines show the $S=32\pm 2$~MeV $M-R$ curves from Figure~\ref{fig:RMS}.  }
\label{fig:RMSMmax}
\end{center}
\end{figure}

\subsection{Maximum Mass}

A viable EOS must be able to support a maximum gravitational mass $M$ that is at least as great as the largest reliably measured neutron-star mass.  For masses, the gold standard is neutron stars in relativistic binaries, for which it is possible to measure post-Keplerian parameters such as the Shapiro delay, pericenter precession, and orbital decay due to the emission of gravitational radiation (see \citealt{2009arXiv0907.3219F} for a good discussion of how these parameters are measured and the governing equations).  The precision with which these masses can be measured, plus the reliability of the underlying theory, makes inferred masses the bedrock of astronomical constraints on the EOS of cold high-density matter.  Particularly notable are the mass measurements $M=1.908\pm 0.016~M_\odot$ for PSR~J1614$-$2230 (original mass measurement in \citealt{2010Natur.467.1081D} and current mass measurement in \citealt{2018ApJ...859...47A}), $M=2.01\pm 0.04~M_\odot$ for PSR~J0348+0432 \citep{2013Sci...340..448A}, and $M=2.14^{+0.10}_{-0.09}~M_\odot$ for PSR~J0740+6620 \citep{2019arXiv190406759T}.  

There are intriguing suggestions of even higher-mass neutron stars.  For example, the ``black widow" system PSR~B1757+20 has an estimated mass of $2.40\pm 0.14~M_\odot$ \citep{2011ApJ...728...95V}, and another black widow system, PSR~1311$-$3430, has an estimated mass of $2.68\pm 0.14~M_\odot$ \citep{2012ApJ...760L..36R}.  However, these measurements are less reliable than the relativistic binary masses because of potential systematic errors and the residuals in the fits \citep{2011ApJ...728...95V,2012ApJ...760L..36R}.  

There are also arguments based on short gamma-ray bursts \citep{2013PhRvL.111m1101B,2015ApJ...812...24F,2015ApJ...808..186L} that were later applied to the double-neutron-star coalescence event GW170817 \citep{2017ApJ...850L..19M}, which suggest a relatively {\it low} maximum mass.  For example, \citet{2017ApJ...850L..19M} suggest that if the two neutron stars in GW170817 formed a hypermassive neutron star that collapsed within tens or hundreds of milliseconds to a black hole, then $M_{\rm max}\ltorder 2.17~M_\odot$, which is precisely consistent with the predictions of \citet{2015ApJ...812...24F} and \citet{2015ApJ...808..186L}.  However, there is no direct evidence that there was a collapse to a black hole.  Similarly, there are various model-dependent upper limits on $M_{\rm max}$ that have been obtained via comparison of simulations with the kilonova that followed GW170817 (e.g., \citealt{2017PhRvD..96l3012S,2018ApJ...852L..25R,2018PhRvD..97b1501R,2019MNRAS.489L..91C}).  

We adopt, as our standard maximum mass constraint, the combination of three mass measurements: $M=2.14^{+0.10}_{-0.09}~M_\odot$ for PSR~J0740+6620, $M=2.01\pm 0.04~M_\odot$ for PSR~J0348+0432 and $M=1.908\pm 0.016~M_\odot$ for PSR~J1614$-$2230.  We also explore the constraints we would obtain if there is a future mass measurement of $M=2.3\pm 0.1~M_\odot$, or a future mass measurement of $M=2.4\pm 0.1~M_\odot$, or a confirmed {\it upper} limit of $M_{\rm max}=2.2\pm 0.05~M_\odot$.  In all cases, we assume that the masses or mass limits have Gaussian likelihoods.

Figure~\ref{fig:EOSSMmax} shows the second level of constraints, in which we consider that the probability distribution of the symmetry energy is a Gaussian with a mean of 32~MeV and a standard deviation of 2~MeV, and add the mass constraints described above.  It is clear from the figure that such measurements place important constraints on the high-density EOS.  Likewise, Figure~\ref{fig:RMSMmax} shows the resulting mass$-$radius constraints.  We see that, as expected, lower limits on the maximum mass push radii to higher values, whereas upper limits push them to lower values.

\begin{figure}
\begin{center}
  \resizebox{1.0\textwidth}{!}{\includegraphics{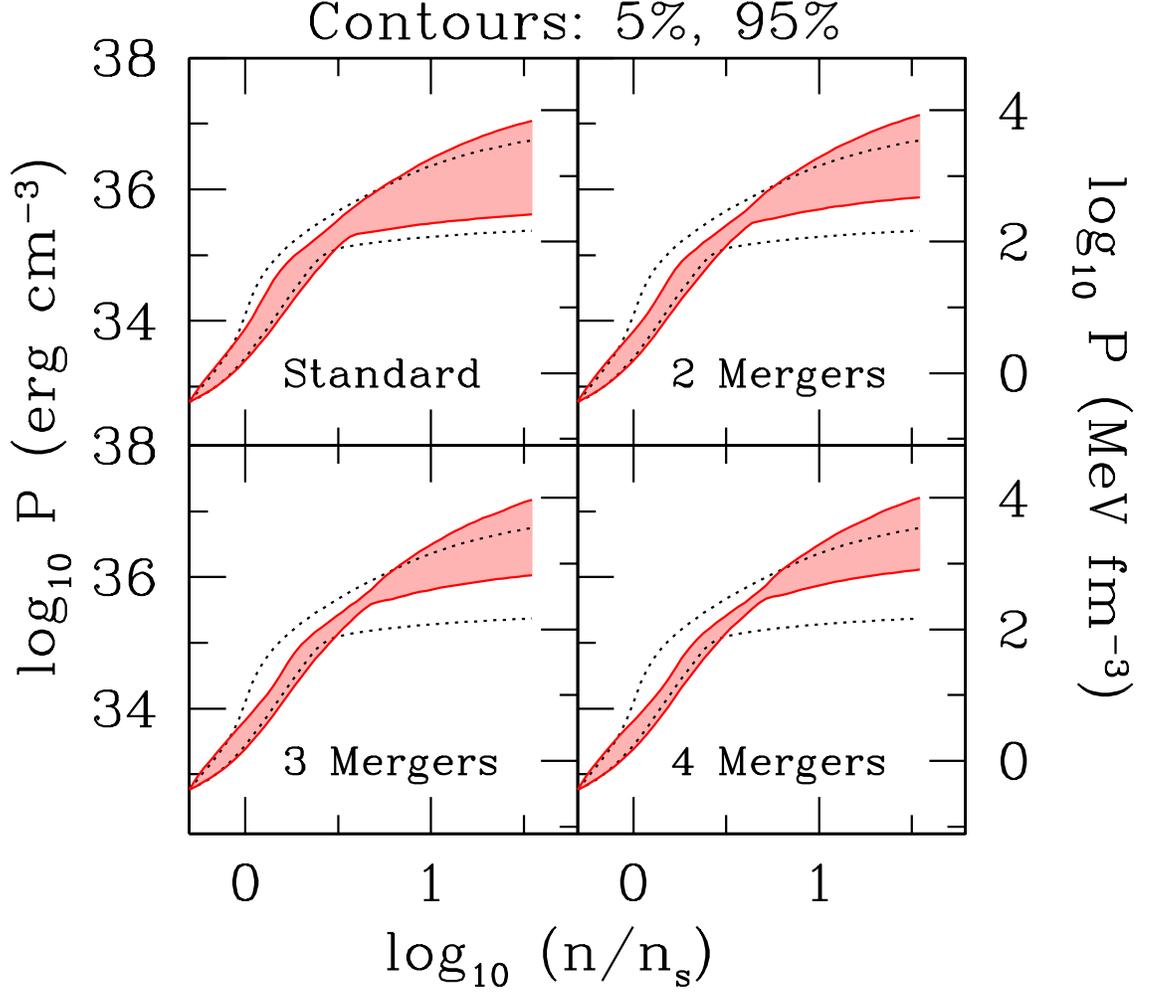}}
\vspace{-2.5truein}
   \caption{Equation-of-state constraints based on symmetry energy, masses, and tidal deformability.  Here we begin with the ``standard" $S+M_{\rm max}$ constraint from Figure~\ref{fig:EOSSMmax}; the dotted lines show the 5\% and 95\% quantiles for that constraint.  The top left panel shows the quantiles when we include constraints based on the tidal deformability of GW170817 \citep{2019PhRvX...9a1001A}.  In order to determine how additional comparable tidal deformability measurements would affect the EOS constraints, in the top right panel we show the consequences of having two events with identical constraints; in the bottom left we suppose we have three events; and in the bottom right we suppose we have four events. Tidal deformability measurements improve our understanding of the EOS at a broad range of densities above nuclear saturation density.  } 
\label{fig:EOSSMmaxL}
\end{center}
\end{figure}

\begin{figure}
\begin{center}
  \resizebox{1.0\textwidth}{!}{\includegraphics{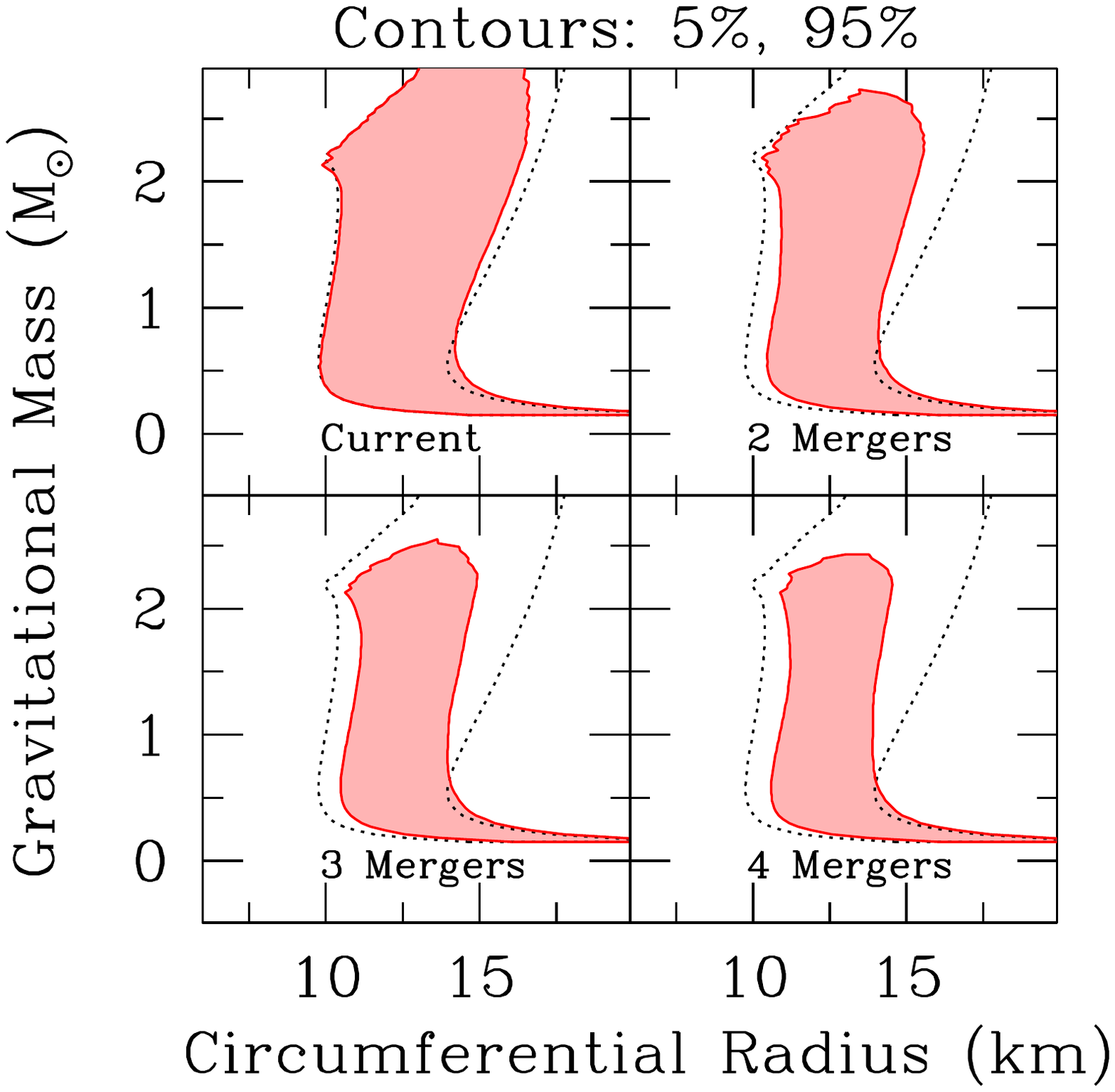}}
\vspace{-2.5truein}
   \caption{Mass$-$radius constraints based on symmetry energy, masses, and tidal deformability.  The constraints shown in each panel correspond to the EOS constraints in the corresponding panel of Figure~\ref{fig:EOSSMmaxL}, and the dotted lines show the 5\% and 95\% quantiles from the ``standard" $S+M_{\rm max}$ $M-R$ curve from Figure~\ref{fig:RMSMmax}.  The relatively low tidal deformability from GW170817 suggests relatively small radii, so if similar constraints are obtained for other events then the preferred radius will decrease.} 
\label{fig:RMSMmaxL}
\end{center}
\end{figure}

\subsection{Tidal Deformability}

The limits on ${\tilde\Lambda}$ from an event such as GW170817 depend on the waveform model, with a spread of $\sim 10$\% among models used thus far \citep{2019PhRvX...9a1001A}.  Bearing this caveat in mind, the middle 90\% of the posterior credible range for ${\tilde\Lambda}$ has been reported as (70,720; \citealt{2019PhRvX...9a1001A}).  Future improvements in gravitational-wave sensitivity, plus the simple accumulation of observing time, are expected to yield a rapidly growing number of detected double-neutron-star coalescences, and potentially a few mergers between neutron stars and black holes.  These additional observations will improve the constraints on the tidal deformability, especially given the anticipated improvements in high-frequency sensitivity due to the use of squeezed light.  It is, however, worth tempering expectations for two reasons: (1)~although tidal effects will be more pronounced at higher frequencies and thus constraints could in principle be improved substantially, waveform families also diverge more at higher frequencies and thus the role of systematic errors will be more prominent, and (2)~GW170817 was an exceptionally strong event (its signal to noise was the largest of any event in the first two LIGO runs; see \citealt{2019PhRvX...9c1040A}),  which means that future events are likely to be measured less precisely.

For GW170817, the full posterior over all model parameters is available at\\ https://dcc.ligo.org/LIGO-P1800061/public.  We use the $\sim 4000$ samples at this site as input for a kernel density estimate (see \citealt{1956AMS...27..832R,1962AMS...33.1065P,1986density.book..S} for details) of the marginalized posterior for the primary mass and binary tidal deformability, which we use in our estimates of the constraints that we display in Figure~\ref{fig:EOSSMmaxL} and Figure~\ref{fig:RMSMmaxL}. Here, we add to our standard $S+M_{\rm max}$ constraints information from tidal deformability measurements.  We begin with the single event GW170817, and then suppose that we have a succession of identical events.  From these figures, it is clear that precise tidal deformability measurements will contribute substantially to our understanding of the dense matter EOS, and to our knowledge of the radius at a wide range of masses.  We also note that various groups have modeled the electromagnetic emission and have proposed other limits on ${\tilde\Lambda}$ (e.g., \citealt{2018ApJ...852L..29R} find a lower limit ${\tilde\Lambda}>400$ for GW170817), but we have not included such limits in our analysis (see \citealt{2019ApJ...876L..31K} for cautionary remarks about lower limits to ${\tilde\Lambda}$ obtained in this manner).

\begin{figure}
\begin{center}
  \resizebox{1.0\textwidth}{!}{\includegraphics{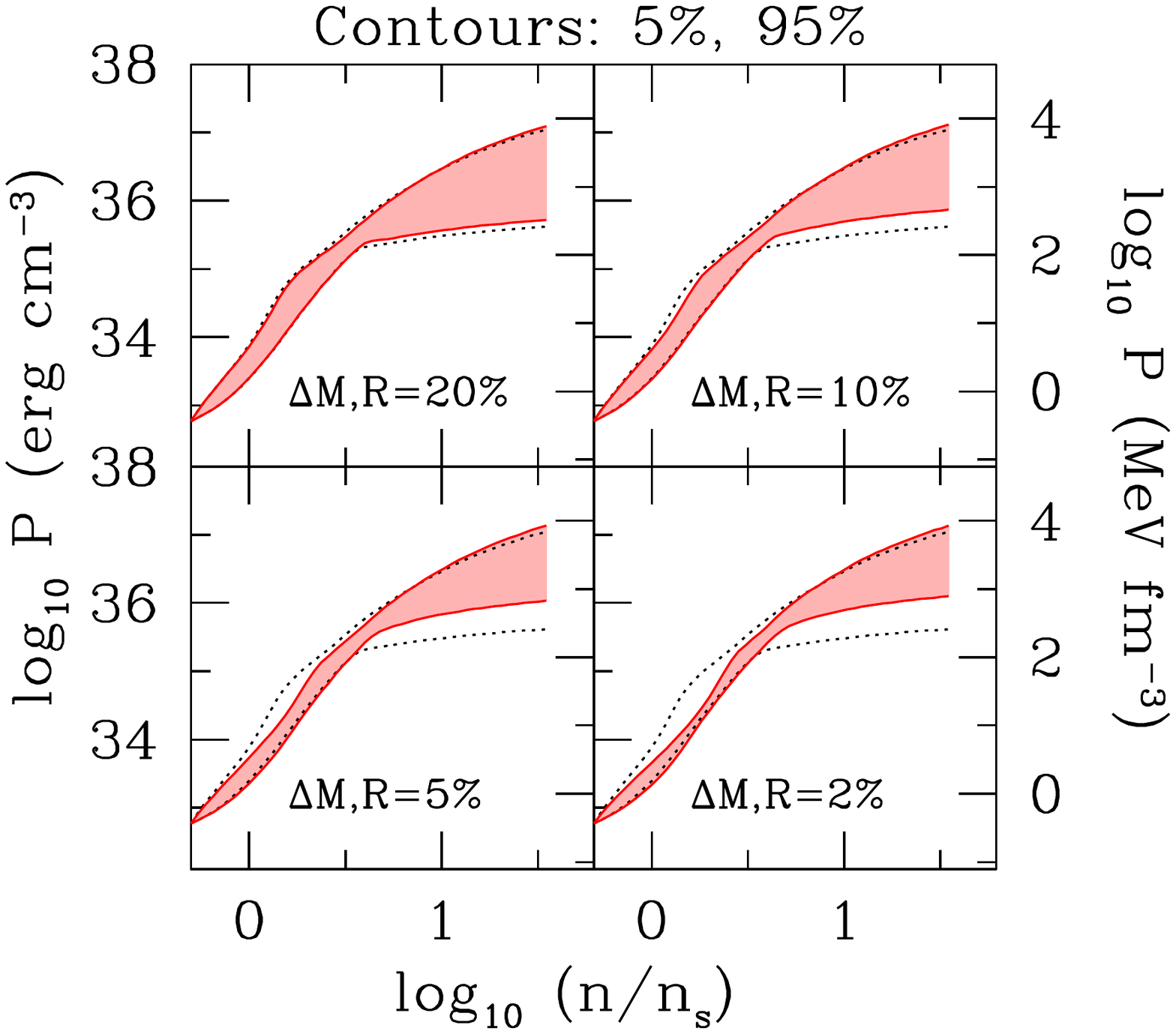}}
\vspace{-2.5truein}
   \caption{Equation-of-state constraints based on measurements of the symmetry energy, masses, and tidal deformability, and illustrative future radius measurements.  Here, we begin with the ``standard" $S+M_{\rm max}+L$ constraint from Figure~\ref{fig:EOSSMmaxL}; the dotted lines show the 5\% and 95\% quantiles for that constraint.  The top left panel shows the effect of adding a measurement of a single $M=1.4~M_\odot$, $R=12$~km star, with fractional Gaussian uncertainties of 20\% for both the mass and radius (see Equation~(\ref{eq:R1})).  The top right panel shows the effect if the fractional Gaussian uncertainties are 10\%, the bottom left assumes uncertainties of 5\%, and the bottom right assumes uncertainties of 2\%.}
\label{fig:EOSSMmaxLR}
\end{center}
\end{figure}

\begin{figure}
\begin{center}
  \resizebox{1.0\textwidth}{!}{\includegraphics{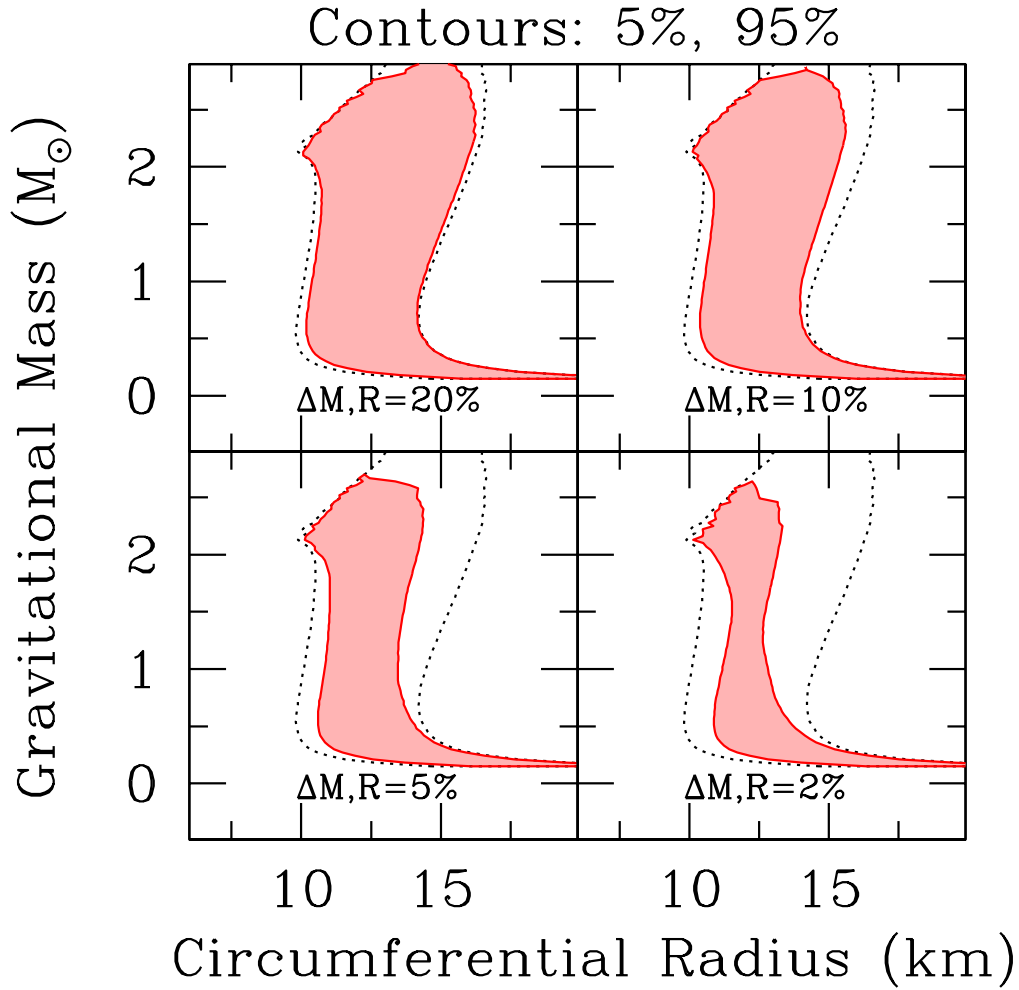}}
\vspace{-2.5truein}
   \caption{Mass$-$radius constraints based on measurements of the symmetry energy, masses, and tidal deformability, and illustrative future radius measurements.  The constraints shown in each panel correspond to the EOS constraints shown in the corresponding panel of in Figure~\ref{fig:EOSSMmaxLR}.  This figure is essentially a check of the algorithm: as should be the case, more precise measurements of mass and radius strongly constrain the mass$-$radius relation.}
\label{fig:RMSMmaxLR}
\end{center}
\end{figure}
\medskip

Thus far, we have used existing measurements, plus plausible extrapolations.  We will now explore the effect of adding additional types of constraints that could be obtained in the future.  

\newpage

\subsection{Radius Measurements}

Reliable and precise radius measurements would be extremely useful in constraining the properties of high-density matter, and much effort has been devoted to the analysis of, in particular, X-ray data from isolated and bursting neutron stars.  However, there are potentially large systematic errors in current reports of neutron-star radii; for detailed discussions, see \citet{2013arXiv1312.0029M} and \citet{2016EPJA...52...63M}, and see additional caveats related to our uncertainty about the EOS of the crust in \citet{2019arXiv190204616G}.  

There is optimism that systematic errors might not be significant for the results that will be obtained using {\it Neutron Star Interior Composition Explorer} ({\it NICER}) measurements of the X-ray pulse waveforms of a few non-accreting neutron stars that are pulsars.  This optimism is based on studies that have been performed of the method, which involves fitting the energy-resolved X-ray waveforms to models with thermally emitting spots that rotate with the neutron star.  \citet{2013ApJ...776...19L} and \citet{2015ApJ...808...31M} generated synthetic waveforms using various geometries and assumptions, and fit them with standard models that had uniformly emitting circular spots.  Although in many cases, the generated spots were oval, or had temperature gradients, or had spectra or beaming patterns different from those assumed in the fitted model, in no case was there a statistically good fit that was significantly biased in mass or radius.  This stands in strong contrast to alternative methods, for which an apparently excellent fit with large bias is possible or even likely, meaning that the fit quality alone does not give a hint that there are potential problems. 

Thus, our opinion is that although current radius measurements may have significant systematic errors, future {\it NICER} measurements are promising.  In addition, as was pointed out by \citet{2018PhRvL.120q2703A} (see also \citealt{2018PhRvL.121i1102D}, \citealt{2018ApJ...857L..23R}, and other papers), gravitational-wave measurements from double-neutron-star mergers can place limits on neutron-star radii, but because these are not independent from tidal deformability estimates, we have not included them separately in our constraints.

In this section, we suppose that a posterior in $(M,R)$ has been obtained for a given star.  The posterior need not be a product of independent posteriors in $M$ and $R$, or independent posteriors in $M/R$ and $M$; the correlations, if any, depend on the details of the system (see \citealt{2013ApJ...776...19L,2015ApJ...808...31M}).  For the purposes of illustration only, we suppose here that the posterior in mass and radius is a product of independent Gaussians:
\begin{equation}
{\cal L}(M,R)\propto e^{-(M-1.4~M_\odot)^2/2\Delta_M^2}e^{-(R-12~{\rm km})^2/2\Delta_R^2}\; ,
\label{eq:R1}
\end{equation}
where we explore the consequences of selecting $\Delta_M$ and $\Delta_R$ equal to 20\%, 10\%, 5\%, and 2\% of the best values of the mass and radius, respectively.

In Figure~\ref{fig:EOSSMmaxLR} we show the effect of adding radius plus mass measurements as described in Equation~(\ref{eq:R1}).  As can be seen in Figure~\ref{fig:EOSSMmaxLR}, a fractional precision of $\ltorder 5$\% for a single star is necessary to add significantly to our information (although as pointed out by \citealt{2016ApJ...822...27M} and \citealt{2019ApJ...881...73W}, in some mass ranges, such as $M>2~M_\odot$, less-precise measurements could still be important).  As a check of our method, we find in Figure~\ref{fig:RMSMmaxLR}, as we must, that improved measurement precision of mass and radius will dramatically tighten the mass$-$radius relation. 

\begin{figure}
\begin{center}
  \resizebox{1.0\textwidth}{!}{\includegraphics{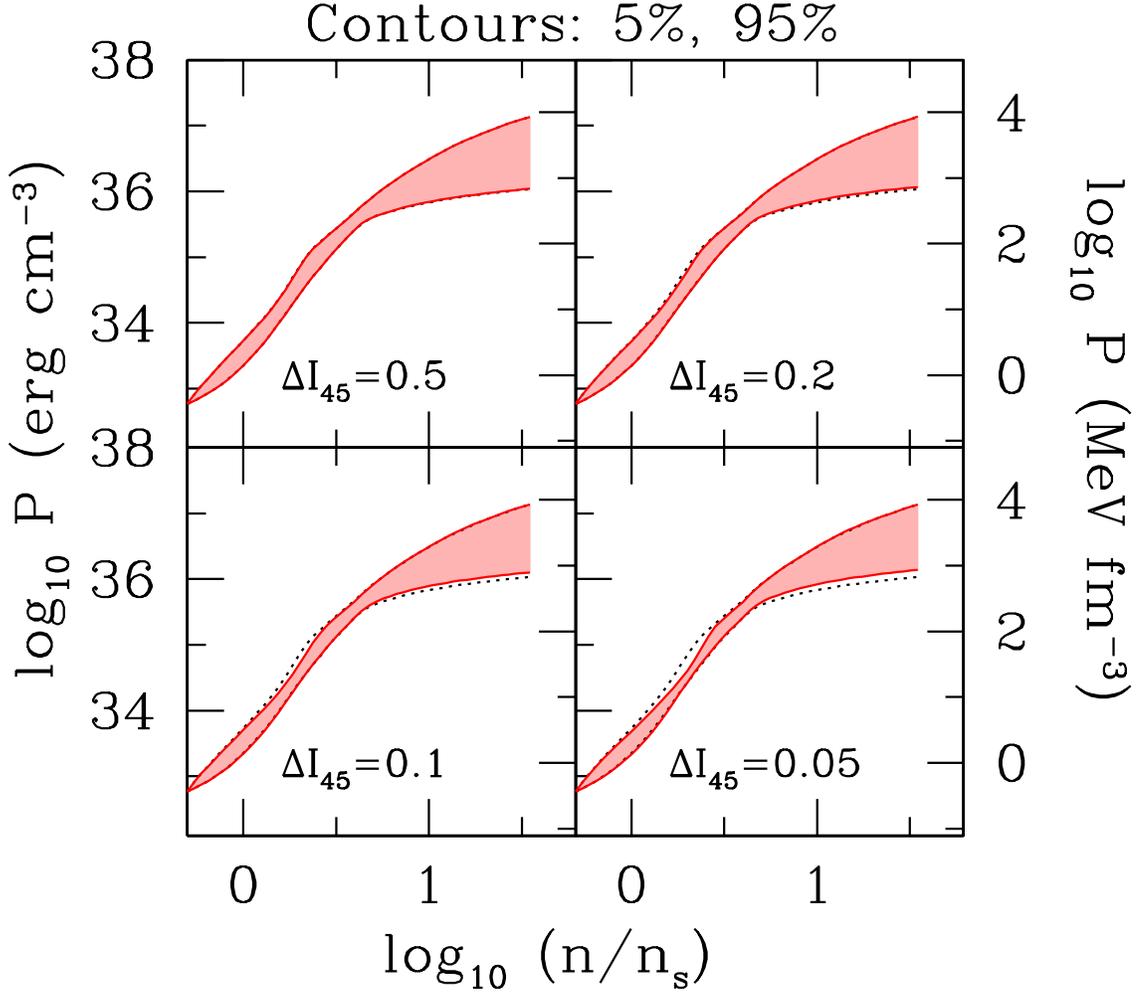}}
\vspace{-2.5truein}
   \caption{Equation-of-state constraints based on measurements of the symmetry energy, masses, and tidal deformability, and illustrative future radius and moment-of-inertia measurements.  Here, we begin with the 5\% $S+M_{\rm max}+L+R$ constraint from Figure~\ref{fig:EOSSMmaxLR}; the dotted lines show the 5\% and 95\% quantiles for that constraint.  The top left panel shows the effect of adding a measurement of the moment of inertia of a $M=1.338~M_\odot$ star, which has a Gaussian distribution centered on $I/10^{45}~{\rm g~cm}^2=1.37$ (the value for an example equation of state with $R=12$~km at $M=1.4~M_\odot$) with a standard deviation of $\Delta I/10^{45}~{\rm g~cm}^2=0.5$.  The top right panel shows the effect of the same measurement with $\Delta I/10^{45}~{\rm g~cm}^2=0.2$, the bottom left panel shows the effect when $\Delta I/10^{45}~{\rm g~cm}^2=0.1$, and the bottom right panel shows the effect when the uncertainty is $\Delta I/10^{45}~{\rm g~cm}^2=0.05$.  Progressively more precise measurements would strongly constrain the EOS at a few times nuclear density.}
\label{fig:EOSSMmaxLRI}
\end{center}
\end{figure}

\begin{figure}
\begin{center}
  \resizebox{1.0\textwidth}{!}{\includegraphics{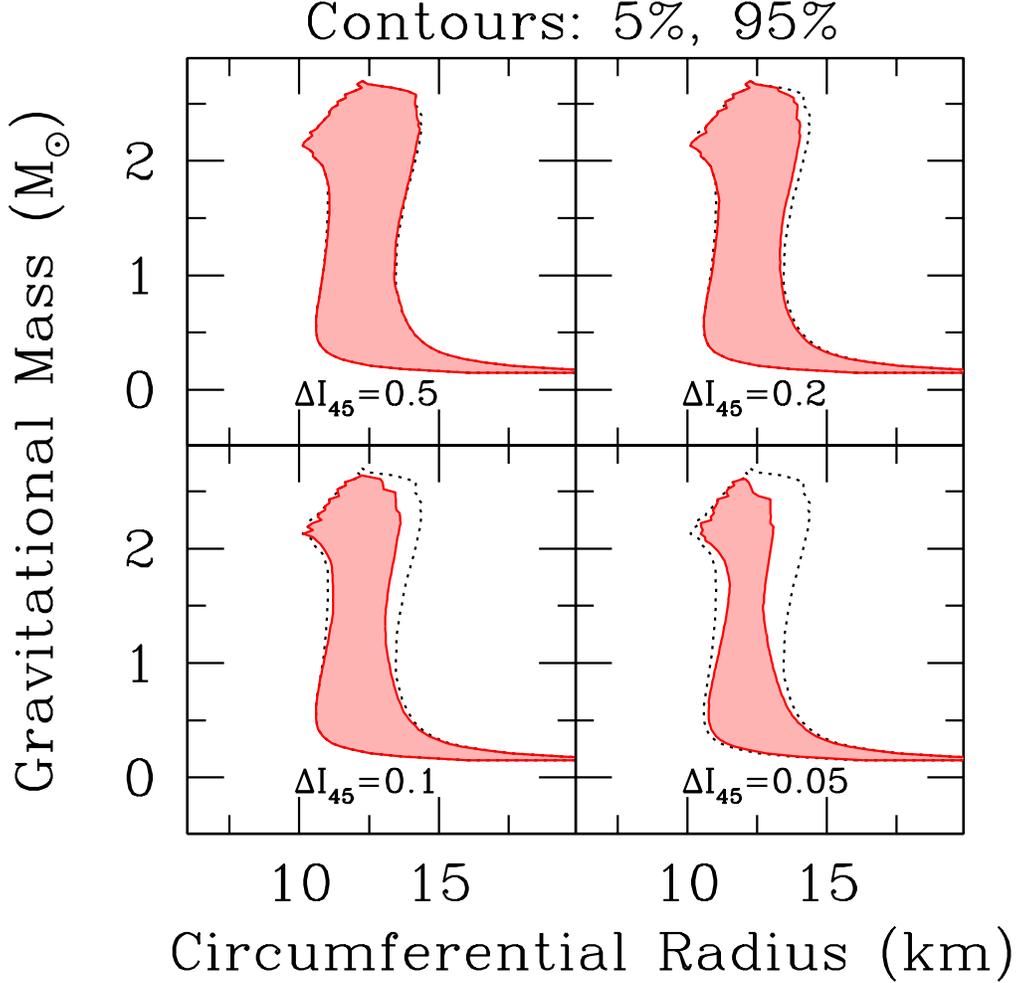}}
\vspace{-2.5truein}
   \caption{Mass$-$radius constraints based on measurements of the symmetry energy, masses, and tidal deformability, and illustrative future radius and moment-of-inertia measurements.  The constraints shown in each panel correspond to the EOS constraints in the corresponding panel of Figure~\ref{fig:EOSSMmaxLRI}.  Improved moment-of-inertia measurements have comparatively small influence on the mass$-$radius relation at low masses, but their influence is significant at $M>1.8~M_\odot$ and for the $I=1.37\times 10^{45}$~g~cm$^2$ that we chose for $M=1.338~M_\odot$, increased precision also reduces the maximum mass.}
\label{fig:RMSMmaxLRI}
\end{center}
\end{figure}

\subsection{Moment of Inertia for Neutron Star of Known Mass}

Shortly after the discovery of the double pulsar PSR~J0737$-$3039 \citep{2003Natur.426..531B}, it was pointed out \citep{2005ApJ...629..979L,2009CQGra..26g3001K} that in principle, spin-orbit coupling could be measured within a few years from the resulting extra pericenter precession, and that this might yield an interestingly precise moment of inertia for the more rapidly rotating of the two pulsars, PSR~J0737$-$3039A (which has a mass of $M=1.3381\pm 0.0007~M_\odot$: \citealt{2006Sci...314...97K}).  The measurement has been far more challenging than originally envisioned, but there is still hope that within about a decade, the moment of inertia can be measured to within $\sim 10$\%.  For our illustrative constraint, we select $I_{1.338}=1.37\times 10^{45}$~g~cm$^2$, with Gaussian uncertainties, because this is consistent with the other real and hypothetical constraints we are applying and because it is consistent with the moment-of-inertia range found by \citet{2018ApJ...868L..22L}.

We add our second hypothetical constraint in Figure~\ref{fig:EOSSMmaxLRI}: moment-of-inertia measurements for the $M=1.338~M_\odot$ neutron star PSR~J0737$-$3039A.  Measurements to the hoped-for precision of $\sim 10$\% for this single star would add significantly to the constraints, but less-precise measurements would have little effect.  In Figure~\ref{fig:RMSMmaxLRI} we see that better moment-of-inertia measurements for an $M=1.338~M_\odot$ star would have little influence on estimates of the radii of stars with masses around $M=1.0~M_\odot$, but would significantly improve the estimates of the radii of $M=1.8-2~M_\odot$ stars.

\begin{figure}
\begin{center}
  \resizebox{1.0\textwidth}{!}{\includegraphics{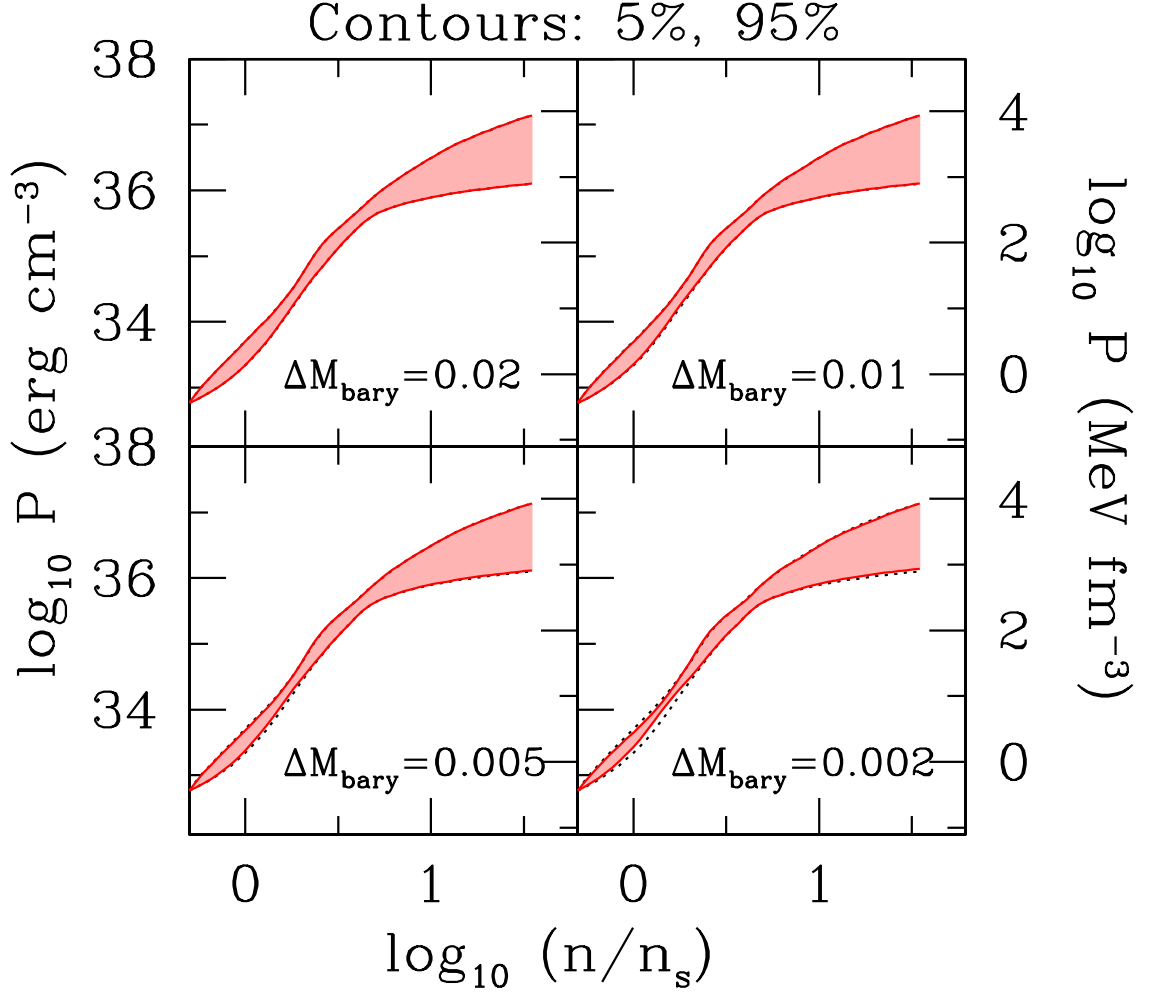}}
\vspace{-2.5truein}
   \caption{Equation-of-state constraints based on measurements of the symmetry energy, masses, and tidal deformability, and illustrative future radius, moment of inertia, and binding energy measurements.  Here, we begin with the 10\% precision $S+M_{\rm max}+L+R+I$ constraint from Figure~\ref{fig:EOSSMmaxLRI}; the dotted lines show the 5\% and 95\% quantiles for that constraint.  The top left panel shows the effect of assuming that for a star with gravitational mass $M=1.2489~M_\odot$ star the probability distribution for the baryonic rest mass is a Gaussian centered on $1.37~M_\odot$ with a standard deviation of $0.2~M_\odot$.  The top right panel assumes a standard deviation of $0.1~M_\odot$, the bottom left $0.05~M_\odot$, and the bottom right $0.02~M_\odot$.  Measurement of the baryonic mass with a precision of of $\sim 0.005~M_\odot$ or better would contribute to our knowledge of the equation of state.}
\label{fig:EOSSMmaxLRIE}
\end{center}
\end{figure}

\begin{figure}
\begin{center}
  \resizebox{1.0\textwidth}{!}{\includegraphics{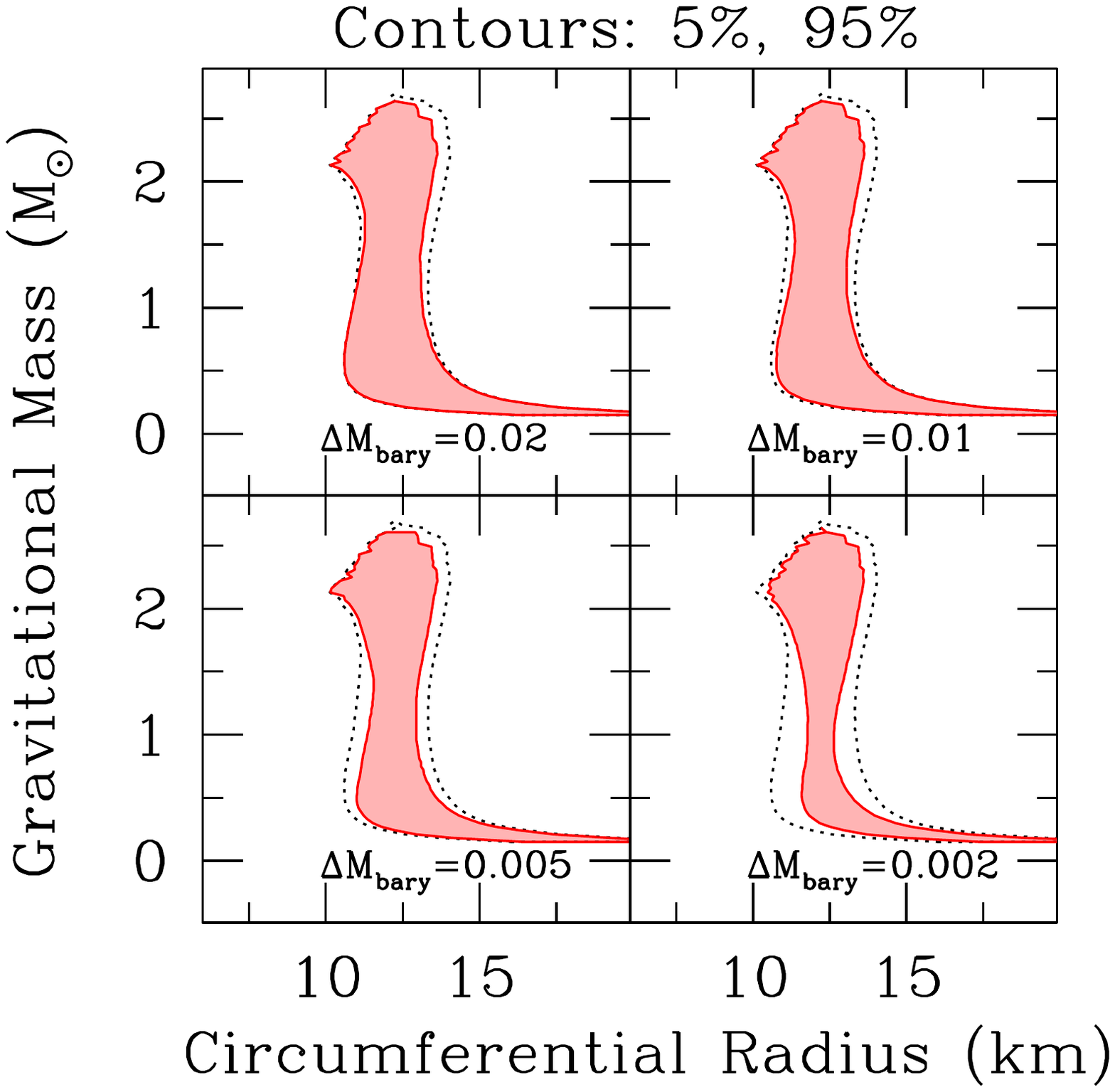}}
\vspace{-2.5truein}
   \caption{Mass$-$radius constraints based on measurements of the symmetry energy, masses, and tidal deformability, and illustrative future radius, moment-of-inertia, and binding energy measurements.  The constraints shown in each panel correspond to the EOS constraints in the corresponding panel of Figure~\ref{fig:EOSSMmaxLRIE}.  The primary influence of a precise binding energy measurement for a low gravitational mass $M=1.2489~M_\odot$ is on the radius at low masses.}
\label{fig:RMSMmaxLRIE}
\end{center}
\end{figure}

\subsection{Binding Energy of Neutron Stars Formed in Electron-capture Supernovae}

If it were possible to know the baryonic rest mass (that is, the sum of the masses of all of the constituent particles if separated to large distance at zero speed) as well as the gravitational mass, for individual neutron stars to reasonable precision, then  the resulting knowledge of the binding energy for those stars would provide another constraint on the EOS.  It is not possible to make a direct measurement of the baryonic rest mass of a star, but there are suggestions that a particular type of core-collapse supernova known as an electron-capture supernova might occur when the core baryonic rest mass is in the narrow range $M_{\rm bary}\sim1.36-1.37~M_\odot$ \citep{1984ApJ...277..791N,2004ApJ...612.1044P,2005MNRAS.361.1243P,2019arXiv190704184Z}.  If there is then neither expulsion of mass nor additional fallback, and if neutron stars formed via this mechanism can be identified and their gravitational masses measured, then a constraint could be applied.  There are clearly several ways in which this identification, or the estimate of the baryonic rest mass, could fail.  Moreover, some objects likely to be neutron stars are too light to have formed from an electron-capture supernova, e.g., the $M=1.174\pm 0.004~M_\odot$ companion to PSR~J0453+1159 \cite{2015ApJ...812..143M}, so other mechanisms to produce low-mass neutron stars (such as ultra-stripped supernovae; see \citealt{2017ApJ...846..170T}) could be in play in this mass range.  Notwithstanding those caveats, \citet{2005MNRAS.361.1243P} made the interesting suggestion that the second pulsar in the double pulsar system, PSR~J0737$-$3039B, originated from an electron-capture supernova, and that its gravitational mass of $M=1.2489\pm 0.0007~M_\odot$ should therefore be identified with $M_{\rm bary}=1.366-1.375~M_\odot$.

Thus, when we incorporate this hypothetical factor into our analysis, we do so by assuming that the baryonic mass corresponding to a gravitational mass $M=1.2489~M_\odot$ is $1.37~M_\odot$ with a Gaussian likelihood.

In Figures~\ref{fig:EOSSMmaxLRIE} and \ref{fig:RMSMmaxLRIE} we show the effect of adding this constraint.  A fractional uncertainty $\ltorder 0.5$\% would improve our knowledge of the EOS below $\sim{\rm few}\times n_s$, and would also tighten the range of radii of low-mass neutron stars.

\subsection{Selected Results for a Piecewise Polytropic Parameterization of the Equation of State}
\label{sec:polytrope}

\begin{figure}
\begin{center}
  \resizebox{1.0\textwidth}{!}{\includegraphics{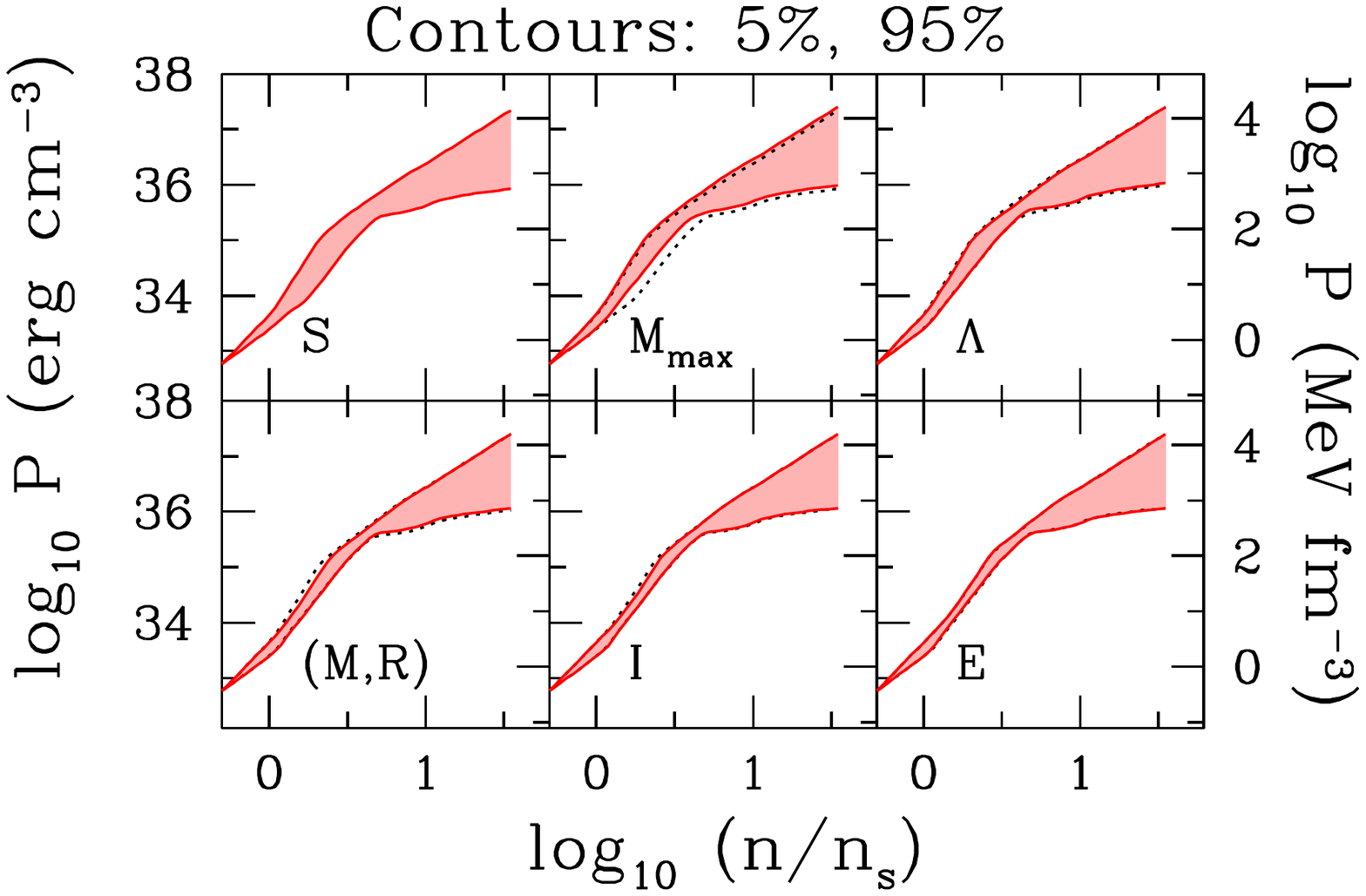}}
\vspace{-3.5truein}
   \caption{Constraints on the equation of state using a parameterization employing a sequence of polytropes (see the text for details).  The line and shading types mean the same as they did for the constraints based on the spectral equation of state.  Here, we use a subset of the real and hypothetical measurements that we discuss above.  We use, sequentially, $S=32\pm 2$~MeV; the masses of the three most massive neutron stars; the tidal deformability of GW170817; a hypothetical $(M,R)=(1.4~M_\odot,12~{\rm km})$ measurement to 5\% precision; a hypothetical measurement of the moment of inertia of a $1.338~M_\odot$ star to 10\% precision; and hypothetical knowledge of the baryonic rest mass of a star to $0.005~M_\odot$ precision.  The constraints are similar, although not identical, to those obtained for the spectral equation of state.}
\label{fig:EOSGammaRho}
\end{center}
\end{figure}

\begin{figure}
\begin{center}
  \resizebox{1.0\textwidth}{!}{\includegraphics{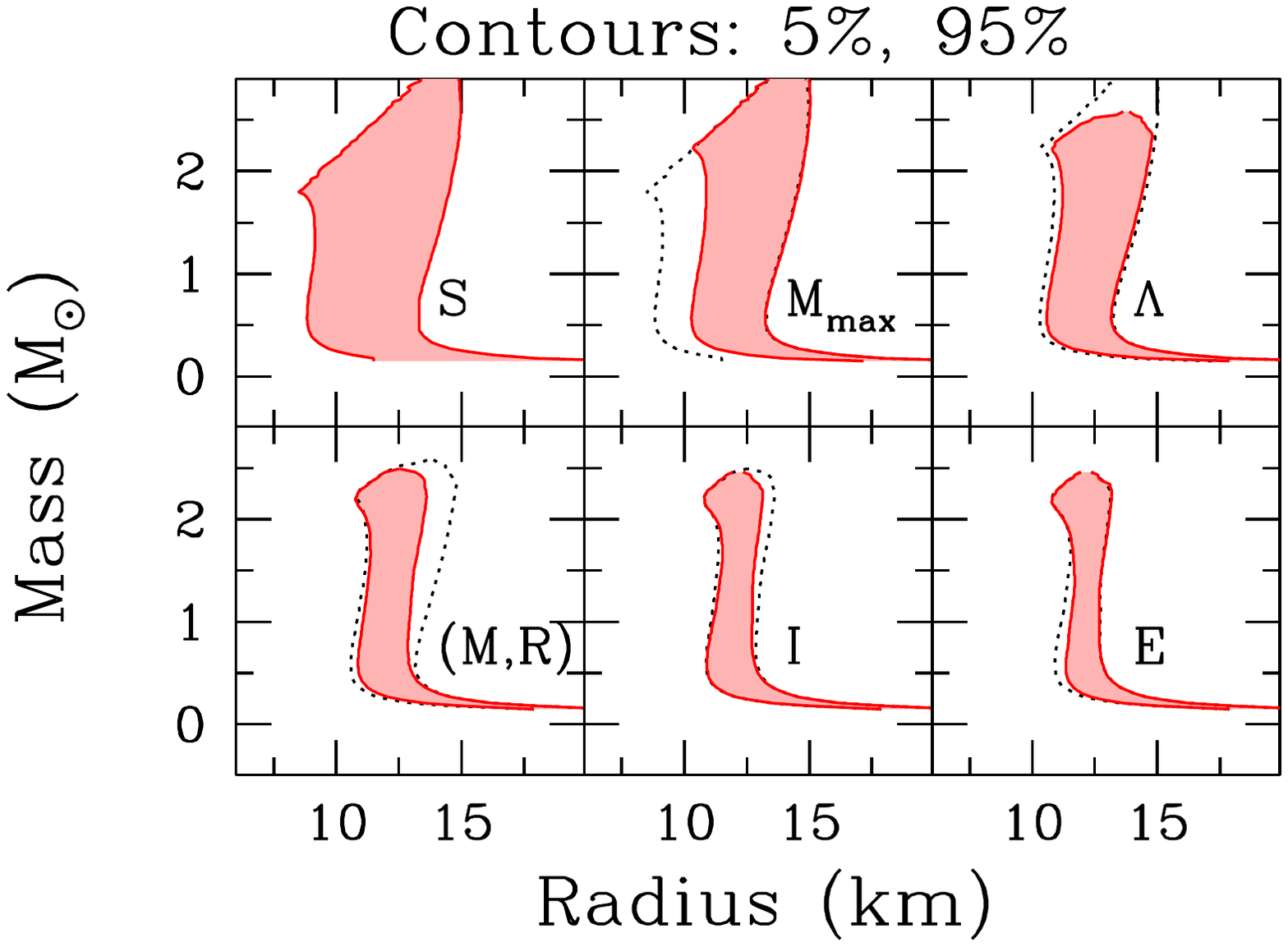}}
\vspace{-3.5truein}
   \caption{Mass$-$radius constraints corresponding to the equation-of-state constraints shown in Figure~\ref{fig:EOSGammaRho}.  Again, the results are quite similar to those we found when we used the spectral parameterization for the equation of state.}
\label{fig:RMGammaRho}
\end{center}
\end{figure}

One drawback of the spectral parameterization we use, with the priors we adopt, is that this does not allow the pressure to be nearly constant over a range of densities.  That is, this parameterization is poor at reproducing phase transitions.  Although we again stress that the main point of our paper is our Bayesian inference framework rather than specific results, we present for comparison results for a subset of the measurements presented above, for a different equation of state.

In this parameterization, we again enforce causality ($dP/d\rho<c^2$) and stability ($dP/d\rho>0$) and use the SLy EOS \citep{2001A&A...380..151D} up to less than half of the nuclear saturation density $\rho_s$.  Above $\rho_0=\rho_s/2$ we represent the EOS by a sequence of polytropes with indices that can change at transition densities that are also parameters: our priors are $\rho_1\in[3/4,5/4]\rho_s$, $\rho_2\in[3/2,5/2]\rho_s$, $\rho_3\in[3,5]\rho_s$, and $\rho_4\in[6,10]\rho_s$.  Our priors on the polytropic indices are $\Gamma_1\in[2,3]$ from $\rho_0$ to $\rho_1$, $\Gamma_2\in[0.1,5]$ from $\rho_1$ to $\rho_2$, $\Gamma_3\in[0.1,5]$ from $\rho_2$ to $\rho_3$, $\Gamma_4\in[0.1,5]$ from $\rho_3$ to $\rho_4$, and $\Gamma_5\in[0.1,5]$ for densities higher than $\rho_4$.  All priors are flat in the bracketed range.  The limited range $[2,3]$ for $\Gamma_1$ is informed by the study of \citet{2013ApJ...773...11H}.

The results of using this parameterization with a subset of our measurements are shown in Figure~\ref{fig:EOSGammaRho} and Figure~\ref{fig:RMGammaRho}.  Here, the progressive measurements are $S=32\pm 2$~MeV; the measured masses of PSR~J0740+6620, PSR~J0348+0432, and PSR~1614$-$2230; the tidal deformability from GW170817; a hypothetical measurement of $(M,R)=(1.4~M_\odot,12~{\rm km})$ with 5\% precision; a hypothetical measurement of the moment of inertia of an $M=1.338~M_\odot$ star with $\Delta I_{45}=0.1$; and hypothetical knowledge to within $0.005~M_\odot$ of the baryonic rest mass of a star with a gravitational mass of $M=1.2489~M_\odot$.  We see that although details of the resulting constraints are somewhat different than for the spectral parameterization, the trends are similar.

\section{Conclusions}
\label{sec:conclusions}

We have shown that diverse sources of both laboratory and astronomical information about cold, dense, catalyzed matter can be incorporated flexibly within a straightforward, rigorous, and practical Bayesian framework.  We treat carefully the constraints that stem from the existing measurements of the symmetry energy, large neutron-star masses, and tidal deformability, the expected future measurements of neutron-star radii and masses, and the possible future measurements of the moments of inertia and gravitational binding energies of neutron stars.  We find that different types of measurements will play significantly different roles in constraining the EOS in different density ranges.  For example, better symmetry energy measurements will have a major influence on our understanding of matter somewhat below nuclear saturation density but little influence above that density.  In contrast, precise radius measurements or multiple tidal deformability measurements of the quality of those from GW170817 or better will improve our knowledge of the equation of state over a broader density range.  Of course, any of these analyses would have to be revisited if systematic errors dominate; but overall, the prospects are good in the next few years for a dramatically enhanced understanding of the nature of dense matter.

\acknowledgements
We thank Chris Pethick for advice on the nuclear physics and Paulo Bedaque, David Blaschke, Reed Essick, Will Farr, Koutarou Kyutoku, Phillipe Landry, and Victor Santos Guedes for other illuminating discussions.  We also thank the anonymous referee for an unusually constructive report.  M.C.M. and C.C. thank the Kavli Institute for Theoretical Physics for its hospitality during the completion of this paper.  M.C.M. and C.C. are grateful for the hospitality of Perimeter Institute where part of this work was carried out. Research at Perimeter Institute is supported in part by the Government of Canada through the Department of Innovation, Science and Economic Development Canada and by the Province of Ontario through the Ministry of Economic Development, Job Creation and Trade. CC was also supported in part by the Simons Foundation through the Simons Foundation Emmy Noether Fellows Program at Perimeter Institute.  This work was partially supported by the Brazilian National Council for Scientific and Technological Development (CMPq).  This research was also supported in part by the National Science Foundation under grant No. NSF PHY-1748958.  The authors acknowledge the use of NASA's Astrophysics Data System (ADS) Bibliographic Services and the arXiv.

\software{SuperMongo (https://www.astro.princeton.edu/$\sim$rhl/sm/sm.html)}

\bibliography{eos}

\end{document}